\pdfoutput=1

\documentclass[reprint,amsmath,amssymb,floatfix,longbibliography,aip,letterpaper,cha]{revtex4-1}

\usepackage{graphics,graphicx,float}
\usepackage{mathtools}		% for underbracket

\makeatletter
\preprintsty@sw{}{%
 \usepackage{titlesec}
 \titleformat{\section}{\normalfont\footnotesize\sffamily\bfseries\uppercase}%
 	{\thesection}{1em}{}%
 \titleformat{\subsubsection}{\normalfont\small\sffamily\slshape}{\thesubsubsection}{1em}{}%
 }%
\makeatother
  
\def\citeyear{\citep}
\def\autocite{\citep}

\usepackage[pdftex,
            pdfauthor={Steven A. Frank},
            pdftitle={How to read probability distributions as statements about process},
            pdfsubject={How to read probability distributions as statements about process}]{hyperref} 

\newcommand*{\Tr}{\text{T}}
\newcommand*{\Trf}{\Tr\mskip-3mu_f}
\newcommand*{\Trfp}{\Tr'\mskip-3mu_f}
\newcommand*{\GR}{\text{G}}
\newcommand*{\En}{{\cal{E}}}
\newcommand*{\surp}{{\cal{S}}}
\newcommand*{\Sy}{\surp_y}

\newcommand*{\Ga}{\alpha}
\newcommand*{\Gb}{\beta}
\newcommand*{\Gd}{\delta}

\newcommand*{\Gg}{\gamma}

\newcommand*{\Gk}{\kappa}
\newcommand*{\Gl}{\lambda}
\newcommand*{\GL}{\Lambda}
\newcommand*{\Gm}{\mu}

\newcommand*{\Gs}{\sigma}

\newcommand*{\Gth}{\theta}
\newcommand*{\Gf}{\phi}
\newcommand*{\Gp}{\psi}

\newcommand*{\dd}{\textrm{d}}

\newcommand*{\EEq}[1]{Eqn~\ref{eq:#1}}
\newcommand*{\Eq}[1]{eqn~\ref{eq:#1}}

\newcommand*{\Eqp}[1]{(Eq.~\ref{eq:#1})}

\newcommand*{\prt}{\partial}

\newcount\BoxNum \BoxNum 1\relax
\makeatletter
\newcommand*{\boxlabel}[1]{%
  \protected@write \@auxout {}{\string \newlabel {box:#1}{{\the\BoxNum}}{}}%
  \advance\BoxNum 1\relax}
\makeatother

\renewcommand*{\EEq}[1]{Eq.~(\ref{eq:#1})}
\renewcommand*{\Eq}[1]{Eq.~(\ref{eq:#1})}

\begin{document}

\title{How to read probability distributions as statements about process}

\author{Steven A.\ Frank}
%\email[email: ]{safrank@uci.edu}
%\homepage[homepage: ]{http://stevefrank.org}
\affiliation{Department of Ecology and Evolutionary Biology, University of California, Irvine, CA 92697--2525  USA}

\begin{abstract}

Probability distributions can be read as simple expressions of information. Each continuous probability distribution describes how information changes with magnitude. Once one learns to read a probability distribution as a measurement scale of information, opportunities arise to understand the processes that generate the commonly observed patterns. Probability expressions may be parsed into four components: the dissipation of all information, except the preservation of average values, taken over the measurement scale that relates changes in observed values to changes in information, and the transformation from the underlying scale on which information dissipates to alternative scales on which probability pattern may be expressed. Information invariances set the commonly observed measurement scales and the relations between them. In particular, a measurement scale for information is defined by its invariance to specific transformations of underlying values into measurable outputs. Essentially all common distributions can be understood within this simple framework of information invariance and measurement scale\footnote{email: \href{mailto:safrank@uci.edu}{safrank@uci.edu}, homepage: \href{http://stevefrank.org}{http://stevefrank.org}}\footnote{\baselineskip=10pt Published as Frank, S. A. 2014. How to read probability distributions as statements about process. \textit{Entropy} \textbf{16:}6059--6098, available at: \href{http://dx.doi.org/10.3390/e16116059}{http://dx.doi.org/10.3390/e16116059}}.  

\bigskip

%\noindent \textbf{Keywords:} measurement; maximum entropy; information theory; statistical mechanics; extreme value distributions.

\end{abstract}

\maketitle

{\renewcommand{\tocname}{}\small\hbox{\null}\vskip-66pt\tableofcontents}

\section{Introduction}

Patterns of nature often follow probability distributions. Physical processes lead to an exponential distribution of energy levels among a collection of particles. Random fluctuations about mean values generate a Gaussian distribution. In biology, the age of cancer onset tends toward a gamma distribution. Economic patterns of income typically match variants of the Pareto distributions with power law tails.

Theories in those different disciplines attempt to fit observed patterns to an underlying generative process. If a generative model predicts the observed pattern, then the fit promotes the plausibility of the model. For example, the gamma distribution for the ages of cancer onset arises from a multistage process \autocite{frank07dynamics}. If cancer requires $k$ different rate-limiting events to occur, then, by classical probability theory, the simplest model for the waiting time for the $k$th event to occur is a gamma distribution. 

Many other aspects of cancer biology tell us that the process indeed depends on multiple events. But how much do we really learn by this inverse problem, in which we start with an observed distribution of outcomes and then try to infer underlying process? How much does an observed distribution by itself constrain the range of underlying generative processes that could have led to that observed pattern?

The main difficulty of the inverse problem has to do with the key properties of commonly observed patterns. The common patterns are almost always those that arise by a wide array of different underlying processes \autocite{jaynes03probability,frank09the-common}. We may say that a common pattern has a wide basin of attraction, in the sense that many different initial starting conditions and processes lead to that same common outcome. For example, the central limit theorem is, in essence, the statement that adding up all sorts of different independent processes often leads to a Gaussian distribution of fluctuations about the mean value. 

In general, the commonly observed patterns are common because they are consistent with so many different underlying processes and initial conditions. The common patterns are therefore particularly difficult with regard to the inverse problem of going from observed distributions to inferences about underlying generative processes. But an observed pattern does provide some information about the underlying generative process, because only certain generative processes lead to the observed outcome. How can we learn to read a mathematical expression of a probability pattern as a statement about the family of underlying processes that may generate it? 

%\nsrpart{Results}

\section{Overview}

In this article, I will explain how to read continuous probability distributions as simple statements about underlying process. I presented the technical background in an earlier article \autocite{frank11a-simple}, with addition details in other publications \autocite{frank09the-common,frank11measurement,frank10measurement}. Here, I focus on developing the intuition that allows one to read probability distributions as simple sentences. I also emphasize key unsolved puzzles in the understanding of commonly observed probability patterns. 

The third section introduces the four components of probability patterns: the dissipation of all information, except the preservation of average values, taken over the measurement scale that relates changes in observed values to changes in information, and the underlying scale on which information dissipates relative to alternative scales on which probability pattern may be expressed.

The fourth section develops an information theory perspective. A distribution can be read as a simple statement about the scaling of information with respect to the magnitude of the observations. Because \textit{measurement} has a natural interpretation in terms of information, we can understand probability distributions as pure expressions of measurement scales.

The fifth section illustrates the scaling of information by the commonly observed log-linear pattern. Information in observations may change logarithmically at small magnitudes and linearly at large magnitudes. The classic gamma distribution is the pure expression of the log-linear scaling of information.

The sixth section presents the inverse linear-log scale. The Lomax and generalized Student's distributions follow that scale. Those distributions include the classic exponential and Gaussian forms in their small-magnitude linear domain, but add power law tails in their large-magnitude logarithmic domain.

The seventh section shows that the commonly observed log-linear and linear-log scales form a dual pair through the Laplace transform. That transform changes addition of random variables into multiplication, and multiplication into addition. Those arithmetic changes explain the transformation between multiplicative log scaling and additive linear scaling. In general, integral transforms describe dualities between pairs of measurement scales, clarifying the relations between commonly observed probability patterns.

The eighth section considers cases in which information dissipates on one scale, but we observe probability pattern on a different scale. The log-normal distribution is a simple example, in which observations arise as products of perturbations. In that case, information dissipates on the additive log scale, leading to a Gaussian pattern on that log scale. 

The eighth section continues with the more interesting case of extreme values, in which one analyzes the largest or smallest value of a sample. For extreme values, dissipation of information happens on the scale of cumulative probabilities, but we express probability pattern on the typical scale for the relative probability at each magnitude. Once one recognizes the change in scale for extreme value distributions, those distributions can easily be read in terms of my four basic components. 

The ninth section returns to dual scales connected by integral transforms. In superstatistics, one evaluates a parameter of a distribution as a random variable rather than a fixed value. Averaging over the distribution of the parameter creates a special kind of integral transform that changes the measurement scale of a distribution, altering that original distribution into another form with a different scaling relation. 

The tenth section considers alternative perspectives on generative process. We may observe pattern on one scale, but the processes that generated that pattern may have arisen on a dual scale. For example, we may observe the classic gamma probability pattern of log-linear scaling, in which we measure the time per event. However, the underlying generative process may have a more natural interpretation on the inverse linear-log scaling of the Lomax distribution. That inverse scale has dimensions of events per unit time, or frequency.

The eleventh section reiterates how to read probability distributions. I then introduce the L\'evy stable distributions, in which dual scales relate to each other by the Fourier integral transform. The L\'evy case connects log scaling in the tails of distributions to constraints in the dual domain on the average of power law expressions. The average of power law expressions describes fractional moments, which associate with the common stretched exponential probability pattern. 

The twelfth section explains the relations between different probability patterns. Because a probability pattern is a pure expression of a measurement scale, the genesis of probability patterns and the relations between them reduce to understanding the origins of measurement scales. The key is that the dissipation of information and maximization of entropy set a particular invariance structure on measurement scales. That invariance strongly influences the commonly observed scales and thus the commonly observed patterns of nature. 

The twelfth section continues by showing that particular aspects of invariance lead to particular patterns. For example, shift invariance with respect the information in underlying values and transformed measured values leads to exponential scaling of information. By contrast, affine invariance leads to linear scaling. The distinctions between broad families of probability distributions turn on this difference between shift and affine invariance for the information in observations.

The thirteenth section presents a broad classification of measurement scales and associated probability patterns. Essentially all commonly observed distributions arise within a simple hierarchically generated sequence of measurement scales. That hierarchy shows one way to consider the genesis of the common distributions and the relations between them. I present a table that illustrates how the commonly observed distributions fit within this scheme. 

The fourteenth section considers the most interesting unsolved puzzle: Why do linear and logarithmic scaling dominate the base scales of the commonly observed patterns? One possibility is that linear and log scaling express absolute and relative incremental information, the two most common ways in which information may scale. Linear and log scaling also have a natural association with addition and multiplication, suggesting a connection between common arithmetic operations and common scaling relations. 

The fifteenth section suggests one potential solution to the puzzle of why commonly observed measurement scales are simple. Underlying values may often be transformed by multiple processes before measurement. Each transformation may be complex, but the aggregate transformation may smooth into a simple relation between initial inputs and final measured outputs. The scaling that defines the associated probability pattern must provide invariant information with respect to underlying values or final measured outputs. If the ultimate transformation of underlying values to final measured outputs is simple, then the required invariance may often define a simple information scaling and associated probability pattern. 

The Discussion summarizes key points and emphasizes the major unsolved problems.

\section{The four components of probability patterns}

To parse probability patterns, one must distinguish four properties. In this section, I begin by briefly describing each property. I then match the properties to the mathematical forms of different probability patterns, allowing one to read probability distributions in terms of the four basic components. Later sections develop the concepts and applications.

First, dissipation of information occurs because most observable phenomena arise by aggregation over many smaller scale processes. The multiple random, small scale fluctuations often erase the information in any particular lower level process, causing the aggregate observable probability pattern to be maximally random subject to constraints that preserve information \autocite{jaynes57information,jaynes57informationII,jaynes03probability}. 

Second, average values tend to be the only preserved information after aggregation has dissipated all else. \textcite{jaynes57information,jaynes57informationII,jaynes03probability} developed dissipation of information and constraint by average values as the key principles of maximum entropy, a widely used approach to understanding probability patterns. I extended Jaynesian maximum entropy by the following components \autocite{frank11a-simple,frank11measurement,frank10measurement}.

Third, average values may arise on different measurement scales. For example, in large scale fluctuations, one might only be able to obtain information about the logarithm of the underlying values. The constrained average would be the mean of the logarithmic values, or the geometric mean. The information in measurements may change with magnitude. In some cases, the scale may be linear for small fluctuations but logarithmic for large fluctuations, leading to an observed linear-log scale of observations. 

Fourth, the measurement scale on which information dissipates may differ from the scale on which one observes pattern. For example, a multiplicative process causes information to dissipate on the additive logarithmic scale, but we may choose to analyze the observed multiplicative pattern. Alternatively, information may dissipate by the multiplication of the cumulative probabilities that individual fluctuations fall below some threshold, but we may choose to analyze the extreme values of aggregates on a transformed linear scale.

The measurement scaling defines the various commonly observed probability distributions. By learning to parse the scaling relations of measurement implicit in the mathematical expressions of probability patterns, one can read those expression as simple statements about underlying process. The previously hidden familial relations between different kinds of probability distributions become apparent through their related forms of measurement scaling.

\subsection{Dissipation of information}

Most observations occur on a macroscopic scale that arises by aggregation of many small scale phenomena \autocite{gibbs02elementary}. Each small scale process often has a random component. The greater the number of small scale fluctuations that combine to form an aggregate, the greater the total randomness in the macroscopic system. We may think of randomness as entropy or as the loss of information. Thus, aggregation dissipates information and increases entropy \autocite{jaynes57information,jaynes57informationII}. 

A typical measure of entropy or randomness is
\begin{equation}\label{eq:entropy}
	\En = -\int p_y \log(p_y) \dd y,
\end{equation}
in which $p_y$ describes the probability distribution for a variable $y$. 

Information is the negative of the entropy, and so the dissipation of information is also given by the entropy \autocite{cover91elements}. I use a continuous form of entropy throughout this article, and focus only on the continuous probability distributions. Discrete distributions follow a similar logic, but require different expressions and details of presentation. 

We can find the probability distribution consistent with maximum entropy by maximizing the expression in \Eq{entropy}, which requires solving $\prt \En/\prt p_y = 0$. The solution is $p_y = c$, where $c$ is a constant. This uniform distribution describes the pattern in which the probability of observing any value is the same for all values of $y$. The maximum entropy uniform distribution has the least information, because all outcomes are equally likely.

\subsection{Constraint by average values}

Suppose that we are studying the distribution of energy levels in a population of particles. We want to know the probability that any particle has a certain level of energy. The probability distribution over the population describes the probability of different levels of energy per particle. 

Typically, there is a certain total amount of energy to be distributed among the particles in the population. The fixed total amount of energy constrains the average energy per particle. 

To find the distribution of energy, we could reasonably assume that many different processes operate at a small scale, influencing each particle in multiple ways. Each small scale process often has a random component. In the aggregate of the entire population, those many small scale random fluctuations tend to increase the total entropy in the population, subject to the constraint that the mean is set extrinsically.

For any pattern influenced by small-scale random fluctuations, the only constraint on randomness may be a given value for the mean. If so, then pattern follows maximum entropy subject to a constraint on the mean \autocite{jaynes57information,jaynes57informationII}.

\subsubsection*{Constraint on the mean}

When we maximize the entropy in \Eq{entropy} to find the probability distribution consistent with the inevitable dissipation of information and increase in entropy, we must also account for the constraint on the average value of observable events. The technical approach to maximizing a quantity, such as entropy, subject to a constraint is the method of Lagrange multipliers. In particular, we must maximize the quantity
\begin{equation}\label{eq:lagrange}
	\GL = \En - \Gk C_0-\Gl C,
\end{equation}
in which the constraint on the average value is written as $C=\int p_y y \dd y - \Gm$. The integral term of the constraint is the average value of $y$ over the distribution $p_y$, and the term, $\Gm$, is the actual average value set by constraint. The method guarantees that we find a distribution, $p_y$, that satisfies the constraint, in particular that the average of the distribution that we find is indeed equal to the given constraint on the average, $\int p_y y \dd y = \Gm$. We must also set the total probability to be one, expressed by the constraint $C_0=\int p_y  \dd y - 1$.

We find the maximum of \Eq{lagrange} by solving $\prt \En/\prt p_y = 0$ for the constants $\Gk$ and $\Gl$ that satisfy the constraint on total probability and the constraint on average value, yielding
\begin{equation}\label{eq:exponential}
 p_y \propto e^{-\Gl y},
\end{equation}
in which $\Gl=1/\Gm$, and $\propto$ means ``is proportional to.'' The total probability over a distribution must be one. If we use that constraint on total probability, we can find $\Gk$ such that $\Gp e^{-\Gl y}$ would be an equality rather than a proportionality for $p_y$ for some constant, $\Gp$. That is easy to do, but adds additional steps and a lot of notational complexity without adding any further insight. I therefore present distributions without the adjusting constants, and write the distributions as ``$p_y \propto $'' to express the absence of the constants and the proportionality of the expression. 

The expression in \Eq{exponential} is known as the exponential distribution, or sometimes the Gibbs or Boltzmann distribution. We can read the distribution as a simple statement. The exponential distribution is the probability pattern for a positive variable that is most random, or has least information, subject to a constraint on the mean. Put another way, the distribution contains information only about the mean, and nothing else.

\subsubsection*{Constraint on the average fluctuations from the mean}

Sometimes we are interested in fluctuations about a mean value or central location. For example, what is the distribution of errors in measurements? How do average values in samples vary around the true mean value? In these cases, we may describe the intrinsic variability by the variance. If we constrain the variance, we are constraining the average squared distance of fluctuations about the mean.

We can find the distribution that is most random subject to a constraint on the variance by using the variance as the constraint in \Eq{lagrange}.  In particular, let $C=\int p_y(y-\Gm)^2\dd y - \Gs^2$, in which $\Gs^2$ is the variance and $\Gm$ is the mean.  This expression constrains the squared distance of fluctuations, $(y-\Gm)^2$, averaged over the probability distribution of fluctuations, $p_y$, to be the given constraint, $\Gs^2$. 

Without loss of generality, we can set $\Gm=0$ and interpret $y$ as a deviation from the mean, which simplifies the constraint to be $C=\int p_y y^2\dd y - \Gs^2$. We can then write the constraint on the mean or the constraint on the variance as a single general expression
\begin{equation}\label{eq:fy}
 C=\int p_y f_y\dd y - \bar{f_y},
\end{equation}
in which $f_y$ is $y$ or $y^2$ for constraints on the mean or variance, respectively, and $\bar{f_y}$ is the extrinsicially set constraint on the mean or variance, respectively. Then the maximization of entropy subject to constraint takes the general form
\begin{equation}\label{eq:general1}
 p_y \propto e^{-\Gl f_y}.
\end{equation}
If we constrain the mean, then $f_y=y$ and $\Gl=1/\Gm$, yielding the exponential form in \Eq{exponential}. If we constrain the variance, then $f_y=y^2$, and $\Gl=1/2\Gs^2$, which is the Gaussian distribution. 

\subsection{The measurement scale for average values}

The constraint on randomness may be transformed by the measurement scale \autocite{frank11a-simple,frank10measurement}. We may write the transformation of the observable values, $f_y$, as $\Tr(f_y)\equiv\Trf$. Here, $f_y$ is $y$ or $y^2$ depending on whether we are interested in the average value or in the average distance from a central location, and $\Tr$ is the measurement scale. Thus, the constraint in \Eq{fy} can be written as
\begin{equation}\label{eq:aveconstraint}
 C=\int p_y \Trf\dd y - \bar{\Tr}\mskip-3mu_f,
\end{equation}
which generalizes the solution in \Eq{general1} to
\begin{equation}\label{eq:general2}
 p_y \propto e^{-\Gl \Trf}.
\end{equation}
This form provides a simple way to express many different probability distributions, by simply choosing $\Trf$ to be a constraint that matches the form of a distribution. For example, the power law distribution, $p_y\propto y^{-\Gl}$, corresponds to the measurement scale $\Trf=\log(y)$. In general, finding the measurement scale and the associated constraint that lead to a particular form for a distribution is useful, because the constraint concisely expresses the information in a probability pattern \autocite{frank11a-simple,frank10measurement}. 

Simply matching probability patterns to their associated measurement scales and constraints leaves open the problem of why particular scalings and constraints arise. What sort of underlying generative processes lead to a particular scaling relation, $\Trf$, and therefore attract to the same probability pattern? I address that crucial question in later sections. For now, it is sufficient to note that we have a simple way to connect the dissipation of information and constraint to probability patterns.

\subsection{The scale on which information dissipates}

In some cases, information dissipates on one scale, but we wish to express the probability pattern on another scale. Suppose that information dissipates on the scale given by $x$, leading to the distribution $p_x$. After obtaining the distribution on the scale $x$ by applying the theory for the dissipation of information and constraint, we may wish to transform the distribution to a different scale, $y$. Here, I briefly mention two distinct types of transformation. Later sections illustrate the crucial role of scale transformations in understanding several important probability patterns. The Methods provides technical details.

\subsubsection*{Change of variable}

The relation between $x$ and $y$ is given by the transformation $x=g(y)$, where $g$ is some function of $y$. For example, we may have $x=\log(y)$. In general, we can use any transformation that has meaning for a particular problem. Several important probability distributions arise by dissipation of information on scales other than the one on which we typically express probability patterns. To understand those distributions, one must recognize the scale on which information dissipates and the transformed scale used to express the probability distribution.

Define $m_y=\left|g'(y)\right|$, where $g'$ is the derivative of $g$ with respect to $y$. The notation $m_y$ emphasizes the term as the measurement scale correction when observing pattern on the scale $y$. Because information dissipates on the scale $x$, we can often find the distribution $p_x$ easily from \Eq{general2}, in which $\Trf$ is a function of $f_x$. Applying the change in measure, $m_y$, we obtain 
\begin{equation}\label{eq:general3}
 p_y \propto m_ye^{-\Gl \Trf}
\end{equation}
in which we replace $f_x$ by the transformed expression $f_{g(y)}$ in the scaling relation $\Trf$. 

The key point is that we have simply made a change of variable from $x$ to $y$. The term $m_y$ adjusts the scaling of the probability pattern for that change of variable.

\subsubsection*{Integral transform}

If we take the average of $e^{-xy}$ over the distribution of $x$, we obtain a new function for each value of $y$, as
\begin{equation*}
 h^*(y) = \int e^{-xy}p_x\dd x,
\end{equation*}
which may be interpreted as a Laplace or Fourier transform of the original distribution, $p_x$. Under some conditions, we can think of the transformed function $h^*(y)$ as a distribution that has a paired relation with the original distribution $p_x$. The transformation creates a pair of related measurement scales that determines the associated pair of probability distributions. We may use other transformation functions besides $e^{-xy}$ to create various pairs of measurement scales and probability distributions. 

\section{Reading probability expressions in terms of measurement and information}

In this section, I show that probability distributions can be read as simple statements about the change in information with the magnitude of the observations. The essential scaling relation $\Trf$ expresses exactly how information changes with magnitude. Because \textit{measurement} has a natural interpretation in terms of information, we can also think of $\Trf$ as an expression of the measurement scale associated with a particular probability distribution.

\subsection{Information and surprise}

The key step arises from interpreting
\begin{equation}\label{eq:surprise}
  \Sy=-\log(p_y)
\end{equation}
in \Eq{entropy} as the translation between probability, $p_y$, and information, $\Sy$. This expression is sometimes called \textit{self-information}, which describes the information in an event $y$ in terms of the probability of that event, $p_y$. I use the symbol $\Sy$ because \textcite{tribus61thermostatics} interpreted this quantity as the surprise associated with the magnitude of the observation, $y$. 

The interpretation of $\Sy$ as surprise arises from the idea that relatively rare events are more surprising. For any initial value of $p_y$, the surprise, $-\log(p_y)=\log (1/p_y)$, increases by $\log(2)$ as $p_y$ decreases by half. Thus, the surprise increases linearly with relative rarity.

Surprise connects to information. If we are surprised by an observation, we learn a lot; if we are not surprised, we had already predicted the outcome to be relatively likely, and we gain little information.

Note that entropy in \Eq{entropy} is equivalent to $\int p_y \Sy\dd y$, which is simply the average amount of surprise over a particular probability distribution. A uniform distribution, in which all values of $y$ are equally likely, has a maximum amount of entropy and a minimum amount of information or surprise. The low surprise occurs because, with any value of $y$ equally likely, we can never be relatively more surprised by observing one particular value of $y$ rather than another. 

\subsection{Scaling relations express the change in information}

The expression $\Sy$ relates information to the magnitude of observations, $y$. We can use that relation to develop an understanding of how information changes with magnitude. The change in information with magnitude captures the essential aspect of \textit{measurement scale.} This notion of information in relation to scale turns out to be the key to understanding probability patterns for continuous variables.

I begin with the general expression for probability patterns in \Eq{general2}, altered slightly  here as
\begin{equation}\label{eq:general2-1}
 p_y = \Gp e^{-\Gl \Trf},
\end{equation}
in which $\Gp$ is a constant that sets the total probability of the distribution to one. In this section, we can ignore the scale transformations and the term $m_y$ that led to \Eq{general3}. Those transformations change the original probability pattern from one scale to another. That change of scale does not alter the relation between information and magnitude on the original scale that determined the form of the probability distribution.

If we take the logarithm of both sides of \Eq{general2-1}, we obtain a general expression for probability patterns in terms of information as 
\begin{equation}\label{eq:affineSy}
  \Sy = \Gp + \Gl \Trf.
\end{equation} 
Thus, the change in information, $\dd\Sy$, compares with the change in the scaling relation for measurement, $\dd\Trf$, as
\begin{equation}\label{eq:infoscale}
 \left|\dd\Sy\right| = \left|\Gl \dd\Trf\right|,
\end{equation}
in which absolute values quantify the magnitude of change. Intuitively, we may think of this expression as the increment of information gained for measuring a change in magnitude on the scale $\Trf$. The parameter $\Gl$ is the relative rate of change of information compared with measured values. 

Note that we can also write
\begin{equation}\label{eq:infoscalepropr}
 \dd\Sy \propto \dd\Trf,
\end{equation}
which means that an increment on the measurement scale is proportional to an increment of information.

\subsection{How to read the exponential and Gaussian distributions}

The exponential distribution in \Eq{exponential} has $\Trf=y$ and $\dd\Sy=\Gl$. The parameter $\Gl=1/\Gm$ is the inverse of the distribution's mean value. The exponential distribution describes a constant increase in information with magnitude, associated with a constant decline in relative probability with magnitude. The rate of increase in information with magnitude is the inverse of the mean.

For the Gaussian distribution with a mean of zero, $\Trf=y^2$ and $\dd\Sy=2\Gl y$. The parameter $\Gl=1/2\Gs^2$ is the inverse of twice the distribution's average squared deviation, leading to $\dd\Sy=y/\Gs^2$. The Gaussian distribution describes a linearly increasing gain (constant acceleration) in information with magnitude, associated with a linearly increasing decline (constant deceleration) in relative probability with magnitude. The rate of the linearly increasing gain in information with magnitude is the inverse of the variance. 

The following sections present the way in which to read a wide variety of common distributions in terms of the scaling relations of information and measurement. Later sections consider the underlying structure and familial relations between commonly observed distributions. That underlying structure arises from the information symmetries that relate different measurements scales to each other.

\section{The log-linear scale}

Cancer incidence illustrates how probability patterns may express simple scaling relations \autocite{frank07dynamics}. For many cancers, the probability $p_y$ that an individual develops disease near the age $y$, among all those born at age zero, is approximately
\begin{equation}\label{eq:gamma}
	p_y \propto y^{k-1} e^{-\Ga y},
\end{equation}
which is the gamma probability pattern. A simple generative model that leads to a gamma pattern is the waiting time for the $k$th event to occur. For example, if cancer developed only after $k$ independent rate-limiting barriers or stages have been passed, then the process of cancer progression would lead to a gamma probability pattern. 

That match between a generative multistage model of process and the observed gamma pattern led many people to conclude that cancer develops by a multistage process of progression. By fitting the particular incidence data to a gamma pattern and estimating the parameter $k$, one could potentially estimate the number of rate-limiting stages required for cancer to develop. Although this simple model does not capture the full complexity of cancer, it does provide the basis for many attempts to connect observed patterns for the age of onset to the underlying generative processes that cause cancer \autocite{frank07dynamics}.

Let us now read the gamma pattern as an expression about the scaling of probability in relation to magnitude. We can then compare the general scaling relation that defines the gamma pattern to the different kinds of processes that may generate a pattern matched to the gamma distribution.

The probability expression in \Eq{gamma} can be divided into two terms. The first term is 
\begin{equation}\label{eq:powerLaw}
y^{k-1} = e^{(k-1)\log(y)},
\end{equation}
which matches our general expression for probability patterns in \Eq{general2} with $\Trf=\log(y)$. This equivalence associates the power law component of the gamma distribution with a logarithmic measurement scale. 

For the second term, $e^{-\Ga y}$, in \Eq{gamma}, we have $\Trf=y$, which expresses linear scaling in $y$. Thus, the two terms in \Eq{gamma} correspond to logarithmic and linear scaling
\begin{equation}\label{eq:loglinear}
	p_y \propto \underbracket[.5pt]{\mskip3mu y^{k-1}}_\text{log} \mskip2mu\times
		\mskip5mu\underbracket[.5pt]{\vphantom{y^{k-1}}\mskip2mu e^{-\Ga y}}_\text{linear},
\end{equation}
which leads to an overall measurement function that has the general log-linear form $\Trf= \log(y)-b y$. For the parameters in this example, $b=\Ga/(k-1)$.

When $y$ is small, $\Trf\approx\log(y)$, and the logarithmic term dominates changes in the information of the probability pattern, $\dd\Sy$, and the measurement scale, $\dd\Trf$. By contrast, when $y$ is large, $\Trf\approx -by$, and the linear term dominates. Thus, the gamma probability pattern is simply the expression of logarithmic scaling at small magnitudes and linear scaling at large magnitudes. The value of $b$ determines the magnitudes at which the different scales dominate.

Generative processes that create log-linear scaling typically correspond to a gamma probability pattern. Consider the classic generative process for the gamma, the waiting time for the $k$th independent event to occur. When the process begins, none of the events has occurred. For all $k$ events to occur in the next time interval, all must happen essentially simultaneously. 

The probability of multiple independent events to occur essentially simultaneously is the product of the probabilities for each event to occur. Multiplication leads to power law expressions and logarithmic scaling. Thus, at small magnitudes, the change in information scales with the change in the logarithm of time. 

By contrast, at large magnitudes, after much time has passed, either the $k$th event has already happened, and the waiting is already over, or $k-1$ events have happened, and we are waiting only for the last event. Because we are waiting for a single event that occurs with equal probability in any time interval, the scaling of information with magnitude is linear. Thus, the classic waiting time problem is a generative model that has log-linear scaling. 

The gamma pattern itself is a pure expression of log-linear scaling. That probability pattern matches any underlying generative process that converges to logarithmic scaling at small magnitudes and linear scaling at large magnitudes. Many processes may be essentially multiplicative at small scales and approximately linear at large scales. All such generative processes will also converge to the gamma probability distribution. In the general case, $k$ is a continuous parameter that influences the magnitudes at which logarithmic or linear scaling dominate.

Later, I will return to this important link between generative process and measurement scale. For now, let us continue to follow the consequences of various scaling relations.

The log-linear scale contains the purely linear and the purely logarithmic as special cases. In \Eq{gamma}, as $k\rightarrow1$, the probability pattern becomes the exponential distribution, the pure expression of linear scaling. Alternatively, as $\Ga\rightarrow0$, the probability pattern approaches the power law form, the pure expression of logarithmic scaling.  

\section{The linear-log scale}

Another commonly observed pattern follows a Lomax or Pareto Type II form
\begin{equation}\label{eq:lomax}
p_y \propto \left(1+\frac{y}{\Ga}\right)^{-k},
\end{equation}
which is associated with the measurement scale $\Trf=\log(1+y/\Ga)$. This distribution describes linear-log scaling. For small values of $y$ relative to $\Ga$, we have $\Trf\rightarrow y/\Ga$, and the distribution becomes
\begin{equation}\label{eq:linearlogexp}
p_y \propto e^{-(k/\Ga) y},
\end{equation}
which is the pure expression of linear scaling. For large values of $y$ relative to $\Ga$, we have $\Trf\rightarrow\log(y/\Ga)$, and the distribution becomes
\begin{equation}\label{eq:linearlogpower}
p_y \propto y^{-k},
\end{equation}
which is the pure expression of logarithmic scaling. 

In these examples, I have used $f_y=y$ in the scaling relation $\Trf=\log(1+f_y/\Ga)$. We can add to the forms of the linear-log scale by using $f_y=(y-\Gm)^2$, describing squared deviations from the mean. To simplify the notation, let $\Gm=0$. Then \Eq{lomax} becomes
\begin{equation}\label{eq:students}
p_y \propto \left(1+\frac{y^2}{\Ga}\right)^{-k},
\end{equation}
which is called the generalized Student's or q-Gaussian distribution \autocite{tsallis09introduction}. When the deviations from the mean are relatively small compared with $\Ga$, linear scaling dominates, and the distribution is Gaussian, $p_y\propto e^{-(k/\Ga) y^2}$. When deviations from the mean are relatively large compared with $\Ga$, logarithmic scaling dominates, causing power law tails, $p_y \propto y^{-2k}$.

\section{Relation between linear-log and log-linear scales}

The specific way in which these two scales relate to each other provides much insight into pattern and process.

\subsection{Common scales and common patterns}

The log-linear and linear-log scales include most of the commonly observed probability patterns. The purely linear exponential and Gaussian distributions arise as special cases. Pure linearity is perhaps rare, because very large or very small values often scale logarithmically. For example, we measure distances in our immediate surroundings on a linear scale, but typically measure very large cosmological distances on a logarithmic scale, leading to a linear-log scaling of distance.

On the linear-log scale, positive variables often follow the Lomax distribution \Eqp{lomax}. The Lomax expresses an exponential distribution with a power law tail. Over a sufficiently wide range of magnitudes, many seemingly exponential distributions may in fact grade into a power law tail, because of the natural tendency for the information at extreme magnitudes to scale logarithmically. Alternatively, many distributions that appear to be power laws may in fact grade into an exponential shape at small magnitudes.

When studying deviations from the mean, the linear-log scale leads to the generalized Student's form. That distribution has a primarily Gaussian shape but with power law tails. The tendency for the tails to grade into a power law may again be the rule when studying pattern over a sufficiently wide range of magnitudes \autocite{tsallis09introduction}.

In some cases, the logarithmic scaling regime occurs at small magnitudes rather than large magnitudes. Those cases of log-linear scaling typically lead to a gamma probability pattern. Many natural observations approximately follow the gamma pattern, which includes the chi-square pattern as a special case.

\subsection{Relations between the scales}

The linear-log and log-linear scales seem to be natural inverses of each other. But what does an inverse scaling mean? We obtain some clues by noting that the mathematical relation between the scales arises from
\begin{equation}\label{eq:laplace}
  \underbracket[.5pt]{\mskip3mu\left(1+\frac{f_y}{\Ga}\right)^{-k}\mskip1mu}_\text{linear-log}
   \propto \mskip4mu
  \int e^{-xf_y}\underbracket[.5pt]{\mskip3mu x^{k-1}e^{-\Ga x}\mskip1mu}_\text{log-linear}\dd x.
\end{equation}
The right side is the Laplace transform of the log-linear gamma pattern in the variable $x$, here interpreted for real-valued $f_y$. That transform inverts the scale to the linear-log form, which is the Lomax distribution for $f_y=y$ or the generalized Student's distribution for $f_y=y^2$. 

This relation between scales is easily understood with regard to mathematical operations \autocite{frank11a-simple,frank10measurement}. The Laplace transform changes the addition of random variables into the multiplication of those variables, and it changes the multiplication of random variables into the addition of those variables \autocite{bracewell00the-fourier}. Logarithmic scaling can be thought of as the expression of multiplicative processes, and linear scaling can be thought of as the expression of additive processes. 

The Laplace transform, by changing multiplication into addition, transforms log scaling into linear scaling, and by changing addition into multiplication, transforms linear scaling into log scaling. Thus, log-linear scaling changes to linear-log scaling. The inverse Laplace transform works in the opposite direction, changing linear-log scaling into log-linear scaling.

The fact that the Laplace transform connects two of the most important scaling relations is interesting. But what does it mean in terms of reading and understanding common probability patterns? The following sections suggest one possibility. 

\section{Dissipation of information on alternative scales}

It may be that information dissipates on one scale, but we observe pattern on a different scale. For example, information may dissipate on the frequency scale of events per unit time, but we may observe pattern on the inverse scale of time per event. Before developing that interpretation of the Laplace pair of inverse scales, it is useful to consider more generally the problem of analyzing pattern on one scale when information dissipates on a different scale.

\subsection{Scale change for data analysis}

Information may dissipate on the scale, $x$, but we may wish to observe or to analyze the data on the transformed scale, $y$. For example, the observations, $y$, may arise by the product of positive random values. Then $x=\log(y)$ would be the sum of the logarithms of those random values. The dissipation of information by the addition of random variables often leads to a Gaussian distribution. By application of \Eq{general2}, we have the distribution of $x=\log(y)$ as 
\begin{equation*}
  p_x\propto e^{-\Gl(x-\Gm)^2},
\end{equation*}
where $\Gm$ is the mean of $x$, and $1/\Gl$ is twice the variance of $x$. On the $x$ scale, the Gaussian distribution has $\Trf=(x-\Gm)^2$. 

Suppose we want the distribution on the scale of the observations, $y$, rather than on the logarithmic scale $x=\log(y)$ on which information dissipates. Then we must apply \Eq{general3} to transform to the scale, $x$, to the scale of interest, $y$, by using $g(y)=\log(y)$, and thus $m_y=g'(y)=y^{-1}$.  Then, from \Eq{general3}, we have the log-normal distribution
\begin{equation*}
  p_y \propto y^{-1}e^{-\Gl\left(\log(y)-\Gm\right)^2},
\end{equation*}
which we match to \Eq{general3} by noting that $m_y=y^{-1}$ and $\Trf=\left(\log(y)-\Gm\right)^2$.

Consider another example, in which information dissipates on the log-linear scale, $x$. By \Eq{general2}, log-linear scaling leads to a gamma distribution
\begin{equation*}
  p_x\propto x^{k-1}e^{-\Ga x},
\end{equation*}
in which the log-linear scale has the form $-\Gl\Tr(x)=(k-1)\log(x)-\Ga x$.

Suppose that we wish to analyze the data on a logarithmic scale, or that we only have access to the logarithms of the observations \autocite{frank11measurement}. Then we must analyze the distribution of $y=\log(x)$, which means that the original scale for the dissipation of information was $x=g(y)=e^y$. Therefore
\begin{equation*}
  -\Gl\Tr\left(g(y)\right)= (k-1)\log(e^y)-\Ga e^y.
\end{equation*} 
Because $m_y=g'(y)=e^y$, by \Eq{general3}, we have
\begin{equation*}
  p_y \propto m_y e^{-\Gl\Tr\left(g(y)\right)}=e^y e^{(k-1)\log(e^y)-\Ga e^y},
\end{equation*}
which simplifies to
\begin{equation}\label{eq:expgamma}
  p_y \propto e^{ky-\Ga e^y}.
\end{equation}
We read this as the dissipation of information on the log-linear scale, $x$, and a change of variable $x=e^y$, in order to analyze the log transformation of the underlying distribution as $y=\log(x)$. Data, such as the distribution of biological species abundances in samples, often have an underlying log-linear structure associated with the gamma distribution. Typically, such data are log-transformed before analysis, leading to the distribution in \Eq{expgamma}, which I call the exponential-gamma distribution \autocite{frank11measurement}.  

\EEq{expgamma} has the same form as the commonly observed Gumbel distribution that arises in extreme value theory. That theory turns out to be another way in which information dissipates on one scale, but we analyze pattern on a different scale.

\subsection{Extreme values: dissipation on the cumulative scale}

Many problems depend only on the largest or smallest value of a sample. Extreme values determine much of the financial risk of disasters, the probability of structural failure, and the expectation of unacceptable traffic congestion. In biology, the most advantageous beneficial mutations may set the pace and extent of adaptation. 

At first glance, it may seem that the most extreme values associated with rare events would be hard to predict. Although it is true that the extreme value in any particular case cannot be guessed with certainty, it turns out that the probability distribution of extreme values often follows a very regular pattern. That regularity of extreme values arises from the same sort of strong convergence by which the central limit theorem leads to the regularity of the Gaussian probability distribution.

I describe how the extreme value distributions can be understood by the dissipation of information and scale transformation. I focus on the largest value in a sample. The same logic applies to the smallest value. I emphasize an intuitive way in which to read the extreme value distributions as expressions about process. 

Many sources provide background on the extreme value distributions \autocite{embrechts97modeling,kotz00extreme,coles01an-introduction,gumbel04statistics}. In my own work, I described the technical details for a maximum entropy interpretation of extreme values \autocite{frank09the-common,frank14generative}, and the scale transformations that connect extreme value forms to general measurement interpretations of probability patterns \autocite{frank11a-simple}.

\subsubsection*{Dissipation of information}

In a sufficiently large sample, the probability of an extreme value depends only on the chance that an observation falls in the upper tail of the underlying distribution from which the observations are drawn. All other information about the underlying process dissipates. The average tail probability of the underlying distribution sets the constraint on retained information, expressed as follows.

Let $x$ be the upper tail probability of a distribution, $p_z$, defined as 
\begin{equation}\label{eq:extremeCum}
  x = \int_y^\infty p_z\dd z,
\end{equation}
in which $y$ is a threshold value, and $x$ is the cumulative probability in the upper tail of the distribution $p_z$ above the value $y$. Thus, $x$ is the probability of observing a value that is greater than $y$. The cumulative probability, $x$, tells us how likely it is to observe a value greater than $y$, and thus how likely it is that $y$ would be near the extreme value in a sample of observations.

On the scale, $x$, the dissipation of information in repeated samples causes the distribution of upper tail probabilities to take on the general form of \Eq{general2}, in particular
\begin{equation*}
  p_x \propto e^{-\Gl x}.
\end{equation*}
The average value of $x$, which is the average upper tail probability, sets the only constraint that shapes the pattern of the distribution. We can, without loss of generality, rescale $x$ so that $\Gl=1$, and thus $p_x$ is proportional to $e^{-x}$. 

\subsubsection*{Scale transformation}

Scale transformation describes how to go from tail probabilities, $x$, to the extreme value in a sample, $y$. Suppose tail probabilities, which are on the scale $x$, are related to extreme values, which are on the scale $y$. The relation between $x$ and $y$ is given by \Eq{extremeCum}. We can express that relation as $x=\Tr(y)=\Trf$, in which $\Trf$ is the right-hand side of \Eq{extremeCum}. 

We can now use our general approach to scale transformation in \Eq{general3}, repeated here
\begin{equation*}
 p_y \propto m_y e^{-\Gl \Trf}.
\end{equation*}
In this case, $m_y=\left|\Trfp\right|$, which is the absolute value of the derivative of $x$ with respect to $y$, yielding
\begin{equation}\label{eq:extremeVal}
 p_y \propto \left|\Trfp\right|e^{-\Gl \Trf}.
\end{equation}
This expression provides the general form of probability distributions when $\Trf$ describes the measurement scale for $y$ in terms of the cumulative distribution, or tail probabilities, for some underlying distribution.

The form of $\Trf$ arises, as always, from the information constrained by an average value. For example, if in \Eq{extremeCum} the tail probability decays exponentially such that $p_z\rightarrow e^{-z}$, then
\begin{equation*}
  x = \int_y^\infty p_z\dd z \approx e^{-y}.
\end{equation*}
The average tail probability is the average of $e^{-y}$, and $x=\Trf=e^{-y}$. From \Eq{extremeVal}, we have
\begin{equation}\label{eq:gumbel}
 p_y \propto e^{-y-\Gl e^{-y}},
\end{equation}
which is the Gumbel form of the extreme value distributions. Alternatively, if the average tail probability is the average of $y^{-\Gg}$, from the tail of an underlying distribution that decays as a power law in $y$, then $\Trf=y^{-\Gg}$, and 
\begin{equation*}
 p_y \propto y^{-(\Gg+1)}e^{-\Gl y^{-\Gg}},
\end{equation*}
which is the Fr\'echet form of the extreme value distributions. In summary, the extreme value distributions follow the simple maximum entropy form. The constraint is the average tail probability of an underlying distribution. We transform from the scale, $x$, of the cumulative distribution of tail probabilities, to the scale, $y$, of the extreme value in a sample \autocite{frank09the-common,frank11a-simple}.

\section{Pairs of alternative scales by integral transform}

The prior section discussed paired scales, in which information dissipates on one scale, but we observe pattern on a transformed scale. In those particular cases, the dual relation between scales was obvious. For example, we may explicitly choose to study pattern by a log or exponential transformation of the observations. Or information may dissipate on the cumulative scale of tail probabilities, but we transform to the scale of observed extreme values to express probability patterns.

I now return to the linear-log and log-linear scales, which lead to the most commonly observed probability patterns. How can we understand the duality between these inverted scales? Is there a general way in which to understand the pairing between inverted measurement scales?

\subsection{Overview}

The distributions based on linear-log and log-linear scales form naturally inverted pairs connected by the Laplace transform. \EEq{laplace} showed that connection, repeated here
\begin{equation}\label{eq:laplace2}
  \underbracket[.5pt]{\mskip3mu\left(1+\frac{f_y}{\Ga}\right)^{-k}\mskip1mu}_\text{linear-log}
   \propto \mskip4mu
  \int e^{-xf_y}\underbracket[.5pt]{\mskip3mu x^{k-1}e^{-\Ga x}\mskip1mu}_\text{log-linear}\dd x.
\end{equation}

In this section, I summarize two ways in which to understand this mathematical expression. First, the pair may arise from superstatistics \autocite{beck03superstatistics}, in which a parameter of a distribution is considered to vary rather than to be fixed. Second, the pair provides an example of a more general way in which dual measurement scales connect to each other through integral transformation, which changes one measurement scale into another. Fourier, Laplace, and superstatistics transformations can be understood as special cases of the more general integral transforms. Those general transforms include as special cases the classic characteristic functions and moment generating functions of probability theory.

The following section considers cases in which information dissipates on one of the scales, but we observe pattern on the inverted scale. This duality provides an essential way in which to connect the scaling and constraints of process on one scale to the patterns of nature that we observe on the dual scale. Reading probability patterns in terms of underlying process may often depend on recognizing this essential duality.

\subsection{Superstatistics}

The transformation between scales in \Eq{laplace2} can be interpreted as averaging over a varying parameter. Assume that we begin with a distribution in the variable $f_y$, given by $\Gf(f_y|x)$. Here, $x$ is the parameter of the distribution. Typically, we think of a parameter $x$ as a fixed constant. Suppose, instead, that $x$ varies according to a distribution, $h(x)$. For example, we may think of a composite population in which $f_y$ varies according to $\Gf(f_y|x)$ in different locations, with the mean of the distribution, $1/x$, varying across locations. 

If we measure the composite population, we study the distribution $\Gf(f_y|x)$ when averaged over the different values of $x$, which vary according to $h(x)$. The composite population then follows the distribution given by
\begin{equation}\label{eq:superstat}
  h^*(f_y) = \int \Gf(f_y|x) h(x)\dd x.
\end{equation}
Averaging a distribution, such as $\Gf$, over a variable parameter, is sometimes called \textit{superstatistics} \autocite{beck03superstatistics}.  When the initial distribution, $\Gf(f_y|x)$, is exponential, $e^{-xf_y}$, then superstatistical averaging over the variable parameter $x$ in \Eq{superstat} is equivalent to the Laplace transform, of which \Eq{laplace2} is an example.

\subsection{Integral transforms} 

We may read \Eq{superstat} as an integral transform, which provides a general relation between a pair of measurement scales. Thus, we may think of \Eq{superstat} as a general way in which to express the duality between paired measurement scales, rather than a specific superstatistics process of averaging over a variable parameter. 

In this general integral transform interpretation, we start with some distribution $h(x)$, which has a scaling relation, $\Tr(x)$. Integrating over the transformation kernel $\Gf(f_y|x)$ creates the distribution $h^*(f_y)$, with scaling relation $\Tr^*(f_y)$.  Thus, averaging over the transformation kernel $\Gf$ changes the variable from $x$ to $f_y$, and changes the measurement scale from $\Tr(x)$ to $\Tr^*(f_y)$. 

The interpretation of such scale transformations depends on the particular transformation kernel, which creates the particular properties of the dual relation. The Laplace transform, with the exponential transformation kernel $e^{-xf_y}$, has many special properties that connect paired measurement scales in interesting ways.

\subsection{Scale inversion by the Laplace transform} 

Suppose the log-linear scaling pattern occurs for the variable $x$, as in \Eq{laplace2}. That equation shows that the Laplace transformation kernel, $e^{-xf_y}$, transforms the log-linear scaling relation of $x$ into the linear-log scaling relation of $f_y$, for real values of $f_y$.

The Laplace change of variable from $x$ to $f_y$ often inverts the dimensional units. The exponent of the transformation kernel $e^{-xf_y}$ is usually dimensionless, which means that the dimensions of $x$ and $f_y$ must cancel. Thus, the units of $f_y$ are typically the inverse of the units of $x$. For example, if $x$ has units of time per event, then $f_y$ has units of events or repetitions per time, which is a kind of frequency. The units may also be changed inversely from frequency to time.

The Laplace transform changes the way in which independent observations combine to produce aggregate pattern. On one scale, the distribution of the sum (convolution) of independent observations from an underlying distribution transforms to multiplication of the distributions on the other scale. Inversely, the distribution of multiplied observations on one scale transforms to addition of variables on the other scale. This duality between addition and multiplication on inverted scales corresponds to the duality between linear and logarithmic measurement on the paired scales.

\section{Alternative descriptions of generative process}

We often wish to associate an observed probability pattern with the underlying generative process. The generative process may dissipate information directly on the measurement scale associated with the observed probability pattern. Or, the generative process may dissipate information on a different scale, but we observe the pattern on a transformed scale.

Consider, as an example, the Laplace duality between the linear-log and log-linear scales in \Eq{laplace2}. Suppose that we observe the gamma pattern of log-linear scaling. We wish to associate that observed gamma pattern to the underlying generative process. 

The generative process may directly create a log-linear scaling pattern. The classic example concerns waiting time for the $k$th independent event. For small times, the $k$ events must happen nearly simultaneously. As noted earlier, the probability of multiple independent events to occur essentially simultaneously is the product of the probabilities for each event to occur. Multiplication leads to power law expressions and logarithmic scaling. Thus, at small magnitudes, the change in information scales with the change in the logarithm of time. 

By contrast, at large magnitudes, after much time has passed, either the $k$th event has already happened, and the waiting is already over, or $k-1$ events have happened, and we are waiting only for the last event. Because we are waiting for a single event that occurs with equal probability in any time interval, the scaling of information with magnitude is linear. Thus, the classic waiting time problem expresses a generative model that has log-linear scaling. 

Any process that scales log-linearly tends to the gamma pattern by the dissipation of all other information. The only requirement is that, in the aggregate, small magnitude events associate with underlying multiplicative combinations of probabilities, and large event magnitudes associate with additive combinations. 

In this case, we move from underlying process to observed pattern: a process tends to scale log-linearly, and dissipation of information on that scale shapes pattern into the gamma distribution form. But often we are concerned with the inverse problem. We observe the log-linear gamma pattern, and we want to know what process caused that pattern.

The duality of the log-linear and linear-log scales in \Eq{laplace2} means that a generative process could occur on the linear-log scale, but we may observe the resulting pattern on the log-linear scale. For example, the number of events per unit time (frequency) may combine in a linear, additive way at small frequencies and in a multiplicative, logarithmic way at large frequencies. That linear-log process would often converge to a Lomax distribution of frequency pattern, or to a Student's distribution if we measure squared deviations, $f_y=y^2$. If we observe the outcome of that process in terms of the inverted units of time per event, those inverted dimensions lead to log-linear scaling and a gamma pattern, or to a gamma pattern with a Gaussian tail if we measure squared deviations.

Is it meaningful to say that the generative process and dissipation of information arise on a linear-log scale of events per unit time, but we observe the pattern on the log-linear scale of time per event? That remains an open question. 

On the one hand, the scaling relations and dissipation of information contain exactly the same information whether on the linear-log or log-linear scales. That equivalence suggests a single underlying generative process that may be thought of in alternative ways. In this case, we may consider constraints on average frequency or, equivalently, constraints on average time. More generally, constraints on either of a dual pair of scales with inverted dimensions would be equivalent.

On the other hand, the meaning of constraint by average value may make sense only on one of the scales. For example, it may be meaningful to consider only the average waiting time for an event to occur. That distinction suggests that we consider the underlying generative process strictly in terms of the log-linear scale. However, if our observations of pattern are confined to the inverse frequency scale, then the observed linear-log scaling would only be a reflection of the true underlying process on the dual log-linear scale. 

All paired scales through integral transformation pose the same issues of duality and interpretation with regard to the connection between generative process and observed pattern.

\section{Reading probability distributions}

In this section, I recap the four components of probability patterns. A clear sense of those four components allows one to read the mathematical expressions of probability distributions as sentences about underlying process.

The four components are: the dissipation of all information; except the preservation of average values; taken over the measurement scale that relates changes in observed values to changes in information; and the transformation from the underlying scale on which information dissipates to alternative scales on which probability pattern may be expressed.

Common probability patterns arise from those four components, described in \Eq{general3} by
\begin{equation}\label{eq:readingGeneral}
  p_y \propto m_y e^{-\Gl\Trf}.
\end{equation}
I show how to read probability distributions in terms of the four components and this general expression. To illustrate the approach, I parse several commonly observed probability patterns. This section mostly repeats earlier results, but does so in an alternative way to emphasize the simplicity of form in common probability expressions.

\subsection{Linear scale}

The exponential and Gaussian are perhaps the most common of all distributions. They have the form
\begin{equation}\label{eq:readingLinear}
  p_y \propto e^{-\Gl f_y}.
\end{equation}
The exponential case, $f_y=y$, corresponds to the preservation of the average value, $\bar{y}$. The Gaussian case, $f_y=(y-\Gm)^2$, preserves the average squared distance from the mean, which is the variance. For convenience, I often set $\Gm=0$ and write $f_y=y^2$ for the squared distance. The exponential and Gaussian express the dissipation of information and preservation of average values on a linear scale. We use either the average value itself or the average squared distance from the mean.

\subsection{Combinations of linear and log scales}

Purely linear scaling is likely to be rare over a sufficiently wide range of magnitudes. For example, one naturally plots geographic distances on a linear scale, but very large cosmological distances on a logarithmic scale. 

On a geographic scale, an increment of an additional meter in distance can be measured directly anywhere on earth. The equivalent measurement information obtained at any geographic distance leads to a linear scale. 

By contrast, the information that we can obtain about meter-scale increments tends to decrease with cosmological distance. The declining measurement information obtained at increasing cosmological distance leads to a logarithmic scale.

The measurement scaling of distances and other quantities may often grade from linear at small magnitudes to logarithmic at large magnitudes. The linear-log scale is given by $\Trf=\log\left(1+f_y/\Ga\right)$. Using that measurement scale in \Eq{readingGeneral}, with $m_y=1$ and $\Gl=k$, we obtain
\begin{equation*}
  p_y \propto \left(1+f_y/\Ga\right)^{-k}.
\end{equation*}
When $f_y$ is small relative to $\Ga$, we get the standard exponential form of linear scaling in \Eq{readingLinear}, which corresponds to the exponential or Gaussian pattern. The tail of the distribution, with $f_y$ greater than $\Ga$, is a power law in proportion to $f_y^{-k}$. An exponential pattern with a power law tail is the Lomax or Pareto type II distribution. A Gaussian with a power law tail is the generalized Student's distribution. 

If one measures observations over a sufficiently wide range of magnitudes, many apparently exponential or Gaussian distributions will likely turn out to have the power law tails of the Lomax or generalized Student's forms. Similarly, observed power law patterns may often turn out to be exponential or Gaussian at small magnitudes, also leading to the Lomax or generalized Student's forms.

Other processes lead to the inverse log-linear scale, which changes logarithmically at small magnitudes and linearly at large magnitudes. The log-linear scale is given by $\Trf=\log(f_y)-bf_y$, in which $b$ determines the transition between log scaling at small magnitudes and linear scaling at large magnitudes. Using that measurement scale in \Eq{readingGeneral} with $m_y=1$ and $f_y=y$, and adjusting the parameters to match earlier notation, we obtain the gamma distribution
\begin{equation*}
  p_y \propto y^{k-1}e^{-\Ga y},
\end{equation*}
which is a power law with logarithmic scaling for small magnitudes and an exponential with linear scaling for large magnitudes. The gamma distribution includes as a special case the widely used chi-square distribution. Thus, the chi-square pattern is a particular instance of log-linear scaling.

If we use the log-linear scale for squared deviations from zero, $f_y=y^2$, then we obtain
\begin{equation*}
  p_y \propto y^{k-1}e^{-\Ga y^2},
\end{equation*}
which is a gamma pattern with a Gaussian tail, expressing log-linear scaling with respect to squared deviations. For $k=2$, this is the well-known Rayleigh distribution.

In some cases, information scales logarithmically at both small and large magnitudes, with linearity dominating at intermediate magnitudes \autocite{frank13input-output}. In a log-linear-log scale, precision at the extremes may depend more strongly on magnitude, or there may be a saturating tendency of process at extremes that causes relative scaling of information with magnitude. Relative scaling corresponds to logarithmic measures. 

Commonly observed log-linear-log patterns often lead to the beta family of distributions \autocite{frank11a-simple}. For example, we can modify the basic linear-log scale, $\Trf=\log\left(1+y/\Ga\right)$, by adding a logarithmic component at small magnitudes, yielding the scale $\Trf=b\log(y)-\log\left(1+y/\Ga\right)$, for $b=\Gg/k$, which leads to a variant of the beta-prime distribution
\begin{equation*}
  p_y \propto y^\Gg\left(1+y/\Ga\right)^{-k}.
\end{equation*}
This distribution can be read as a linear-log Lomax distribution, $\left(1+y/\Ga\right)^{-k}$, with an additional log scale power law component, $y^\Gg$, that dominates at small magnitudes. Other forms of log-linear-log scaling often lead to variants from the beta family.

\subsection{Direct change of scale}

In many cases, process dissipates information and preserves average values on one scale, but we observe or analyze data on a different scale. When the scale change arises by simple substitution of one variable for another, the form of the probability distribution is easy to read if one directly recognizes the scale of change. Here, I repeat my earlier discussion for the way in which one reads the commonly observed log-normal distribution. Other direct scale changes follow this same approach.

If process causes information to dissipate on a scale $x$, preserving only the average squared distance from the mean (the variance), then $x$ tends to follow the Gaussian pattern
\begin{equation*}
 p_x \propto e^{-\Gl(x-\Gm)^2},
\end{equation*}
in which the mean of $x$ is $\Gm$, and the variance is $1/2\Gl$. If the scale, $x$, on which information dissipates is logarithmic, but we observe or analyze data on a linear scale, $y$, then $x=\log(y)$. The value of $m_y$ in \Eq{general3} is the change in $x$ with respect to $y$, yielding $\dd\log(y)/\dd y = y^{-1}$. Thus, the distribution on the $y$ scale is
\begin{equation*}
 p_y \propto y^{-1}e^{-\Gl(\log(y)-\Gm)^2},
\end{equation*}
which is simply the Gaussian pattern for $\log(y)$, corrected by $m_y=y^{-1}$ to account for the fact that dissipation of information and constraint of average value are happening on the logarithmic scale, $\log(y)$, but we are analyzing pattern on the linear scale of $y$. Other direct changes of scale can be read in this way.

\subsection{Extreme values and exponential scaling}

Extreme values arise from the probability of observing a magnitude beyond some threshold. Probabilities beyond a threshold depend on the cumulative probability of all values beyond the cutoff. For an initially linear scale with $f_x=x$, cumulative tail probabilities typically follow the generic form $e^{-\Gl x}$ or, simplifying by using $\Gl=1$, the exponential form $e^{-x}$. The cumulative tail probabilities above a threshold, $y$, define the scaling relation between $x$ and $y$, as
\begin{equation*}
 x = \int^\infty_y e^{-z}\dd z = e^{-y}.
\end{equation*}
Thus, extreme values that depend on tail probabilities tend to define an exponential scaling, $x=e^{-y}=\Trf$. Because we have changed the scale from the cumulative probabilities, $x$, to the probability of some threshold, $y$, that determines the extreme value observed, we must account for that change of scale by $m_y = \left|\Trfp\right|=e^{-y}$, where the prime is the derivative with respect to $y$. Using \Eq{general3} for the generic method of direct change in scale, and using the form of $m_y$ here for the change from the cumulative scale of tail probabilities to the direct scaling of threshold values, we obtain the general form of the extreme value distributions as
\begin{equation*}
 p_y \propto \left|\Trfp\right| e^{-\Gl \Trf}.
\end{equation*}
In this simple case, $\Trf=e^{-y}$, thus
\begin{equation*}
 p_y \propto e^{-y - \Gl e^{-y}},
\end{equation*}
a form of the Gumbel extreme value distribution. Note that this form is just a direct change from linear to exponential scaling, $x=e^{-y}$. 

Alternatively, we can obtain the same Gumbel form by any process that leads to exponential-linear scaling of the form $\Gl\Tr(y) = y + \Gl e^{-y}$, in which the exponential term dominates for small values and the linear term dominates for large values. That scaling leads directly to the distribution
\begin{equation*}
 p_y \propto  \underbracket[.5pt]{\mskip3mu e^{-\Gl e^{-y}}\mskip3mu}_\text{exp} \underbracket[.5pt]{\mskip3mu e^{-y}\mskip3mu}_\text{linear}.
\end{equation*}
The probability of a small value being the largest extreme value decreases exponentially in $y$, leading to the double exponential term $e^{-\Gl e^{-y}}$ dominating the probability. By contrast, the probability of observing large extreme values decreases linearly in $y$, leading to the exponential term $e^{- y}$ dominating the probability.

\subsection{Integral transform and change of scale}

\EEq{laplace} showed the connection between linear-log and log-linear scales through the Laplace integral transform.  The Laplace transform can often be thought of as inverting the dimensional units. For example, we may change from the time per event for a gamma distribution with log-linear scaling to the number of events per unit time (frequency) according to a Lomax distribution with linear-log scaling. Or we may start with a gamma distribution of frequencies and transform to a Lomax distribution of time per event. The units do not have to be in terms of time and frequency. Any pair of inverted dimensions relates to each other in the same way.

That connection between different scales helps to read probability distributions in relation to underlying process. For example, an observation of frequencies distributed according to the linear-log Lomax pattern may suggest dissipation of information and constraint of average values in the dual log-linear measurement domain.

Scale inversion by the Laplace transform also has the interesting property of switching between addition and multiplication in the two domains. For example, multiplicative aggregation of processes and a logarithmic pattern at small magnitudes on the scale of time per event transform to additive aggregation and a linear pattern at small magnitudes on the frequency scale of events per unit time. 

This arithmetic duality of measurement scales clarifies the meaning of probability distributions with respect to underlying generative mechanisms. It would be interesting to study pairs of scales connected by the general integral transform \Eqp{superstat} with respect to the interpretation of aggregation and pattern in dual domains. 

\subsection{L\'evy stable distributions}

Another important family of common distributions arises by a similar scaling duality
\begin{equation}\label{eq:cauchy}
 \left(1+\frac{y^2}{\varphi^2}\right)^{-1} \propto \mskip4mu\int e^{-xiy}\mskip3mu 
 e^{-\varphi|x|}\mskip2mu\dd x.
\end{equation}
Consider each part in relation to the Laplace pair in \Eq{laplace}. The left side is the Cauchy distribution, a special case of the linear-log generalized Student's distribution with $k=1$ and $\Ga=\varphi^2$. On the right, $e^{-\varphi|x|}$ is a symmetric exponential distribution, because $e^{-\varphi x}$ is the classic exponential distribution for $x>0$, and $e^{\varphi x}$ for $x<0$ is the same distribution reflected about the $x=0$ axis. The two distributions together form a new distribution over all positive and negative values of $x$. 

Each positive and negative part of the symmetric exponential, by itself, expresses linearity in $x$. However, the sharp switch in direction and the break in smoothness at $x=0$ induces a quasi-logarithmic scaling at small magnitudes, which corresponds to the linearity at small magnitudes in the transformed domain of the Cauchy distribution.

In this case, the integral transform is Fourier rather than Laplace, using the transformation kernel $e^{-xiy}$ over all positive and negative values of $x$. For our purposes, we can consider the consequences of the Laplace and Fourier transforms as similar with regard to inverting the dimensions and scaling relations between a pair of measurement scales.

The Cauchy distribution is a particularly important probability pattern. In one simple generative model, the Cauchy arises by the same sort of summing up of random perturbations and dissipation of information that leads to the Gaussian distribution by the central limit theorem. The Cauchy differs from the Gaussian because the underlying random perturbations follow logarithmic scaling at large magnitudes. 

Log scaling at large magnitudes causes power law tails, in which the distributions of the underlying random perturbations tend to have the form $1/|x|^{1+\Gg}$ at large magnitudes of $x$. When the tail of a distribution has that form, then the total probability in the tail above magnitudes of $|x|$ is approximately $1/|x|^{\Gg}$. The Cauchy is the particular distribution with $\Gg=1$. Thus, one way to generative a Cauchy is to sum up random perturbations and constrain the average total probability in the tail to be $1/|x|$. 

Note that the constraint on the average tail probability of $1/|x|$ for the Cauchy distribution on the left side of \Eq{cauchy} corresponds, in the dual domain on the right side of that equation, to $e^{-\varphi|x|}$, in which the measurement scale is $\Trf=|x|$. The average of the scaling $\Trf$ corresponds to the preserved average constraint after the dissipation of information. In this case, the dual domain preserves only the average of $|x|$. Thus the dual scaling domains preserve the average of $|x|$ in the symmetric exponential domain and the average total tail probability of $1/|x|$ in the dual Cauchy domain.

We can express a more general duality that includes the Cauchy as a special case by
\begin{equation}\label{eq:levy}
 p_y \propto \mskip4mu\int e^{-xiy}\mskip3mu e^{-\varphi|x|^\Gg}\mskip2mu\dd x.
\end{equation}
The only difference from \Eq{cauchy} is that in the symmetric exponential, I have written $|x|^\Gg$. The parameter $\Gg$ creates a power law scaling $\Trf=|x|^\Gg$, which corresponds to a distribution that is sometimes called a stretched exponential. 

The distribution in the dual domain, $p_y$, is a form of the L\'evy stable distribution. That distribution does not have a mathematical expression that can be written explicitly. The L\'evy stable distribution, $p_y$, can be generated by dissipating all information by summation of random perturbations while constraining the average of the total tail probability to be $1/|x|^\Gg$ for $\Gg<2$. For $\Gg=1$, we obtain the Cauchy distribution. When $\Gg=2$, the distributions in both domains become Gaussian, which is the only case that domains paired by Laplace or Fourier transform inversion have the same distribution. 

Note that the paired scales in \Eq{levy} match a constraint on the average of $|x|^\Gg$ with an inverse constraint on the average tail probability, $1/|x|^\Gg$. Here, $\Gg$ is not necessarily an integer, so the average of $|x|^\Gg$ can be thought of as a fractional moment in the stretched exponential domain that pairs with the power law tail in the inverse L\'evy domain \autocite{frank09the-common}. 

\section{Relations between probability patterns}

I have shown how to read probability distributions as statements about the dissipation of information, the constraint on average values, and the scaling relations of information and measurement. Essentially all common distributions have the form given in \Eq{general3} as
\begin{equation}\label{eq:general3relate}
 p_y \propto m_ye^{-\Gl \Trf}.
\end{equation}
Dissipation of information and constraint on average values set the $e^{-\Gl f_y}$ form. Scaling measures transform the observables, $f_y$, to $\Trf\equiv\Tr(f_y)$. The term $m_y$ accounts for changes between dissipation of information on one scale and measurement of final pattern on a different scale.

The scaling measures, $\Trf$, determine the differences between probability patterns. In this section, I discuss the scaling measures in more detail. What defines a scaling relation? Why are certain common scaling measures widely observed? How are the different scaling measures connected to each other to form families of related probability distributions? 

\subsection{Invariance and common scales}

The form of the maximum entropy distributions influences the commonly observed scales and associated probability distributions \autocite{frank11a-simple,frank10measurement}. In particular, we obtain the same distribution in \Eq{general3relate} for either the measurement function $\Trf$ or the affine transformed measurement function $\Trf \mapsto a+b\Trf$. An affine transformation shifts the variable by the constant $a$ and multiplies it by the constant $b$. 

The shift by $a$ changes the constant of proportionality
\begin{equation*}
  e^{-\Gl(a+\Trf)} = \xi e^{-\Gl\Trf},
\end{equation*}
in which $\xi=e^{-\Gl a}$. In maximum entropy, the final proportionality constant always adjusts to satisfy the constraint that the total probability is one \Eqp{lagrange}. Thus, the final adjustment of total probability erases any prior multiplication of the distribution by a constant. A shift transformation of $\Trf$ does not change the associated probability pattern.

Multiplication by $b$ also has no effect on probability pattern, because
\begin{equation*}
  e^{-\Gl b\Trf} = e^{-\hat{\Gl}\Trf}
\end{equation*}
for $\hat{\Gl}=b\Gl$. In maximum entropy, the final value of the constant multiplier for $\Trf$ always adjusts so that that the average value of $\Trf$ satisfies an extrinsic constraint, as given in \Eq{aveconstraint}. 

Thus, maximum entropy distributions are invariant to affine transformations of the measurement scale. That affine invariance shapes the form of the common measurement scales. In particular, consider transformations of the observables, $\GR(f_y)$, such that
\begin{equation}\label{eq:affine}
  \Tr\left[\GR(f_y)\right] = a + b\Tr(f_y).
\end{equation}
Any scale, $\Tr$, that satisfies this relation causes the transformed scale $\Tr\left[\GR(f_y)\right]$ to yield the same maximum entropy probability distribution as the original scale $\Trf\equiv\Tr(f_y)$.

For example, suppose our only information about a probability distribution is that its form is invariant to a transformation of the observable values $f_y$ by a process that changes $f_y$ to $\GR(f_y)$. Then it must be that the scaling relation of the measurement function $\Trf$ satisfies the invariance in \Eq{affine}. By evaluating how that invariance sets a constraint on $\Trf$, we can find the form of the probability distribution.

The classic example concerns the invariance of logarithmic scaling to power law transformation \autocite{hand04measurement}. Let $\Tr(y) = \log(y)$ and $\GR(y) = c y^\Gg$. Then by \Eq{affine}, we have
\begin{equation}\label{eq:plawinv}
  \log(c y^\Gg) = \log(c) + \Gg\log(y),
\end{equation}
which demonstrates that logarithmic scaling is affine invariant to power law transformations of the form $c y^\Gg$, in which \textit{affine invariance} means that the scaling relation $\Tr$ and the associated transformation $\GR$ satisfy \Eq{affine}. 

\subsection{Affine invariance of measurement scaling}

Put another way, a scaling relation, $\Tr$, is defined by the transformations, $\GR$, that leave unchanged the information in the observables with respect to probability patterns. In maximum entropy distributions, \textit{unchanged} means affine invariance. This affine invariance of measurement scaling in probability distributions is so important that I like to write the key expression in \Eq{affine} in a more compact and memorable form
\begin{equation}\label{eq:affine2}
  \Tr \sim \Tr\circ\GR.
\end{equation}
Here, the circle means composition of functions, such that $\Tr\circ\GR\equiv\Tr[\GR(f_y)]$, and the symbol ``$\sim$'' for similarity means equivalent with respect to affine transformation. Thus, the right side of \Eq{affine} is similar to $\Tr$ with respect to affine transformation, and the left side \Eq{affine} is equivalent to $\Tr\circ\GR$. Reversing sides of \Eq{affine} and using ``$\sim$'' for affine similarity leads to \Eq{affine2}.

Note, from \Eq{affineSy} and \Eq{affine}, that $\Sy\equiv\Tr\circ\GR$, showing that the information in a probability distribution, $\Sy$, is invariant to affine transformation of $\Tr$. Thus, we can also write
\begin{equation*}
  \Tr \sim \Tr\circ\GR \sim \Sy,
\end{equation*}
which emphasizes the fundamental role of invariant information in defining the measurement scaling, $\Tr$, and the associated form of probability patterns.

\subsection{Base scales and notation}

Earlier, I defined $f_y=f(y)$ as an arbitrary function of the variable of interest, $y$. I have used either $y$ or $y^2$ or $(y-\Gm)^2$ for $f_y$ to match the classical maximum entropy interpretation of average values constraining either the mean or the variance. 

To express other changes in the underlying variable, $y$, I introduced the measurement functions or scaling relations, $\Trf\equiv\Tr(f_y)$. In this section, I use an expanded notation to reveal the structure of the invariances that set the forms of scaling relations and probability distributions \autocite{frank11a-simple}. In particular, let 
\begin{equation*}
  w\equiv w(f_y)
\end{equation*}
be a function of $f_y$. Then, for example, we can write an exponential scaling relation as $\Tr(f_y)=e^{\Gb w}$. We may choose a base scale, $w$, such as a linear base scale, $w(f_y)=f_y$, or a logarithmic base scale, $w(f_y)=\log(f_y)$, or a linear-log base scale, $w(f_y)=\log(1+f_y/\Ga)$, or any other base scale. Typically, simple combinations of linear and log scaling suffice. Why such simple combinations suffice is an essential unanswered question, which I discuss later.

Previously, I have referred to $f_y$ as the observable, in which we are interested in the distribution of $y$ but only collect statistics on the function $f_y$. Now, we will consider $w\equiv w(f_y)$ as the observable. We may, for example, be limited to collecting data on $w=\log(f_y)$ or on measurement functions $\Tr(f_y)$ that can be expressed as functions of the base scale $w$. We can always revert to the simpler case in which $w\equiv f_y$ or $w\equiv y$.

In the following sections, the expanded notation reveals how affine invariance sets the structure of scaling relations and probability patterns.

\subsection{Two distinct affine relations}

All maximum entropy distributions satisfy the affine relation in \Eq{affine}, expressed compactly in \Eq{affine2}. In that general affine relation, any measurement function, $\Tr$, could arise, associated with its dual transformation, $\GR$, to which $\Tr$ is affine invariant. That general affine relation does not set any constraints which measurement functions $\Tr$ may occur, although the general affine relation may favor certain scaling relations to be relatively common.

By contrast with the general affine form $\Tr \sim \Tr\circ\GR$, for any $\Tr$ and its associated $\GR$, we may consider how specific forms of $\GR$ determine the scaling, $\Tr$. Put another way, if we require that a probability pattern be invariant to transformations of the observables by a particular $\GR$, what does that tell us about the form of the associated scaling relation, $\Tr$, and the consequent probability pattern?

Here we must be careful about potential confusion. It turns out that an affine form of $\GR$ is itself important, in which, for example, $\GR(w)=\Gd + \Gth w$. That specific affine choice for $\GR$ is distinct from the general affine form of \Eq{affine2}. With that in mind, the following sections explore the consequences of an affine transformation, $\GR$, or a shift transformation, which is a special case of an affine transformation.

\subsection{Shift invariance and generalized exponential measurement scales}

Suppose we know only that the information in probability patterns does not change when the observables undergo shift transformation, such that $\GR(w)=\Gd+w$. In other words, the form of the measurement scale, $\Tr$, must be affine invariant to adding a constant to the base values, $w$. A shift transformation is a special case of an affine transformation $\GR(w)=\Gd+\Gth w$, in which the affine transform becomes strictly a shift transformation for the restricted case of $\Gth=1$.

The exponential scale
\begin{equation}\label{eq:wexp}
  \Trf=e^{\Gb w}
\end{equation}
maintains the affine invariance in \Eq{affine} to a shift transformation, $\GR$. If we apply shift transformation to the observables, $w\mapsto\Gd+w$, then the exponential scale becomes $e^{\Gb (\Gd+w)}$, which is equivalent to $be^{\Gb w}$ for $b=e^{\Gb\Gd}$. We can ignore the constant multiplier, $b$, thus, the exponential scale is shift invariant with respect to \Eq{affine}.

Using the shift invariant exponential form for $\Trf$, the maximum entropy distributions in \Eq{general3relate} become
\begin{equation}\label{eq:expscale}
 p_y \propto m_ye^{-\Gl e^{\Gb w}}.
\end{equation}
This exponential scaling has a simple interpretation. Consider the example in which $w$ is a linear measure of time, $y$, and $\Gb$ is a rate of exponential growth (or decay). Then the measurement scale, $\Trf$, transforms each underlying time value, $y$, into a final observable value after exponential growth, $e^{\Gb y}$. The random time values, $y$, become random values of final magnitudes, such as random population sizes after exponential growth for a random time period. In general, exponential growth or decay is shift invariant, because it expresses a constant rate of change independently of the starting point.

If the only information we have about a scaling relation is that the associated probability pattern is shift invariant to transformation of observables, then exponential scaling provides a likely measurement function, and the probability distribution may often take the form of \Eq{expscale}. 

The Gumbel extreme value distribution in \Eq{gumbel} follows exponential scaling. In that case, the underlying observations, $y$, are transformed into cumulative exponential tail probabilities that, in aggregate, determine the probability that an observation is the extreme value of a sample. The exponential tail probabilities are shift invariant, in the sense that a shifted observation, $\Gd+y$, also yields an exponential tail probability. The magnitude of the cumulative tail probability changes with a shift, but the exponential form does not change.

\subsection{Affine duality and linear scaling}

Suppose probability patterns do not change when observables undergo an affine transformation $\GR(w)=\Gd+\Gth w$. Affine transformation of observables allows a broader range of changes than does shift transformation. The broader the range of allowable transformations of observables, $\GR$, the fewer the measurement functions, $\Tr$, that will satisfy the affine invariance in \Eq{affine}. Thus affine transformation of observables leads to a narrower range of compatible measurement functions than does shift transformation.

When $\GR$ is affine with $\Gth\ne1$, then the associated measurement function $\Trf$ must itself be affine. Because $\Trf$ is invariant to shift and multiplication, we can say that invariance to affine $\GR$ means that $\Trf=w$, and thus the maximum entropy probability distribution in \Eq{general3relate} becomes linear in the base measurement scale, $w$, as
\begin{equation}\label{eq:wlinear}
  p_y\propto m_y e^{-\Gl w}.
\end{equation}
This form follows when the probability pattern is invariant to affine transformation of the observables, $w$. By contrast, invariance to a shift transformation of the observables leads to the broader class of distributions in \Eq{expscale}, of which \Eq{wlinear} is special case for the  more restrictive condition of invariance to affine transformation of observables.

To understand the relation between affine and shift transformations of observables, $\GR$, it is useful to write the expression for the measurement function in \Eq{wexp} more generally as
\begin{equation}\label{eq:tfaffine}
 \Trf = \frac{1}{\Gb}\left(e^{\Gb w}-1\right),
\end{equation}
noting that we can make any affine transformation of a measurement function, $\Trf\mapsto a+b\Trf$, without changing the associated probability distribution. With this new measurement function for shift invariance, as $\Gb\rightarrow0$, then $\Trf\rightarrow w$, and we recover the measurement function associated with affine $\GR$.

Suppose, for example, that we interpret $\Gb$ as a rate of exponential change in the underlying observable, $w$, before the final measurement. Then, as $\Gb\rightarrow0$, the underlying observable and the final measurement become equivalent, $\Trf\rightarrow w$, because
\begin{equation*}
  \Trf=\lim_{\Gb \to 0}\left[\frac{1}{\Gb}\left(e^{\Gb w}-1\right)\right] \rightarrow w.
\end{equation*}

\subsection{Exponential and Gaussian distributions arise from affine invariance}

Suppose we know only that the information in probability patterns does not change when the observables undergo affine transformation, $w\mapsto\Gd+\Gth w$. The invariance of probability pattern to affine transformation of observables leads to distributions of the form in \Eq{wlinear}. Thus, if the observable is the underlying value, $w\equiv y$, then the probability distribution is exponential
\begin{equation*}
  p_y\propto e^{-\Gl y},
\end{equation*}
and if the observable is $y^2$, the squared distance of the underlying value from its mean, then the probability distribution is Gaussian
\begin{equation*}
  p_y\propto e^{-\Gl y^2}.
\end{equation*}
By contrast, if the probability pattern is invariant to a shift of the observables, but not to an affine transformation of the observables, then the distribution falls into the broader class based on exponential measurement functions in \Eq{expscale}.

\section{Hierarchical families of measurement scales and distributions}

The general form for probability distributions in \Eq{expscale}, repeated here
\begin{equation*}
 p_y \propto m_ye^{-\Gl e^{\Gb w}}
\end{equation*}
arises from a base measurement scale, $w$, and shift invariance of the probability pattern to changes $w\mapsto\Gd+w$. Each base scale, $w$, defines a family of related probability distributions, including the linear form
\begin{equation*}
 p_y \propto m_ye^{-\Gl w}
\end{equation*}
as a special case when the probability pattern is invariant to affine changes $w\mapsto\Gd+\Gth w$, which corresponds to $\Gb\rightarrow0$ in \Eq{tfaffine}.

We may consider a variety of base scales, $w$, creating a variety of distinct measurement scales and families of distributions. Ultimately, we must consider how the base scales arise. However, it is useful first to study the commonly observed base scales. The relations between these common base scales form a hierarchical pattern of measurement scales and probability distributions \autocite{frank11a-simple}. 

 \begin{table*}[!t]
  \caption{Some Common Probability Distributions$^*$}
  \label{tab:commonDistn}
  \scriptsize{
  \begin{center}
  \begin{tabular}{|l|l|l|l|}
    \hline
    Distribution & 
    \hfil$p_y$\hfil &
    \hfil$w$\hfil & 
    Notes and alternative names
    \\
    \hline &&&\\[-9pt]
     Gumbel & 
    $e^{\Gb y - \Gl e^{\Gb y}}$ &
    $\text{Linear}$ &
     $m_y=\Tr'$
    \\
   Gibbs/Exponential & 
    $e^{-\Gl y}$ &
   $\text{Linear}$ &
    $\Gb\rightarrow0$
    \\
    Gauss/Normal & 
    $e^{-\Gl y^2}$ &
    $\text{Linear}$ &
     $\Gb\rightarrow0$; $f_y=y^2$
    \\
    Rayleigh & 
    $ye^{-\Gl y^2}$ &
    $\text{Linear}$ &
     $\Gb\rightarrow0$; $f_y=y^2$; $m_y=\Tr'$
    \\
    Log-Normal & 
    $y^{-1}e^{-\Gl \left(\log y\right)^2}$ &
    $\text{Linear}$ &
    $\Gb\rightarrow0$; $f_y=y^2$; $y\rightarrow\log y$; $m_y=y^{-1}$
    \\
    Stretched exponential & 
    $e^{-\Gl y^\Gb}$ &
    $\text{Log}^{(1)}$ &
    Gauss with $\Gb = 2$
   \\
    Fr\'echet/Weibull& 
    $y^{\Gb-1}e^{-\Gl y^\Gb}$ &
    $\text{Log}^{(1)}$ &
    $m_y=\Tr'$; Rayleigh with $\Gb=2$
    \\
    Symmetric L\'evy & 
    $e^{-\Gl |y|^\Gb}$ (Fourier domain) &
    $\text{Log}^{(1)}$  &
    $f_y=|y|$; $\Gb \le 2$; Gauss ($\Gb=2$), Cauchy ($\Gb=1$); \Eq{levy}
    \\
    Pareto type I& 
    $y^{-\Gl}$ &
    $\text{Log}^{(1)}$ &
    $\Gb\rightarrow0$; $m_y=1$ or $m_y=\Tr'$
    \\
    Log-Fr\'echet& 
    $y^{-1}(\log y)^{\Gb-1}e^{-\Gl(\log y)^\Gb}$ &
    $\text{Log}^{(2)}$ &
    $m_y=\Tr$; also from Fr\'echet: $y\rightarrow\log y$, $m_y=y^{-1}\Tr'(y)$
    \\
    ?? & 
    $e^{-\Gl(\log y)^\Gb}$ &
    $\text{Log}^{(2)}$ &
    Also from stretched exponential with $f_y=\log y$
    \\
    Log-Pareto type I & 
    $y^{-1}\left(\log y\right)^{-\Gl}$ &
    $\text{Log}^{(2)}$&
    $\Gb\rightarrow0$; $m_y=\Tr'$; also from Pareto I: $y\rightarrow\log y$, $m_y=y^{-1}$
    \\
    ?? & 
    $\left(\log y\right)^{-\Gl}$ &
    $\text{Log}^{(2)}$ &
    $\Gb\rightarrow0$; also from Pareto I with $f_y=\log y$
    \\
    Pareto type II/Lomax & 
    $\left(c_1+y\right)^{-\Gl}$ &
    $\text{LinLog}^{(1)}$ &
    $\Gb\rightarrow0$
    \\
    Generalized Student's & 
    $\left(c_1+y^2\right)^{-\Gl}$ &
    $\text{LinLog}^{(1)}$ &
    $\Gb\rightarrow0$; $f_y=y^2$; Pearson VII, Kappa; Cauchy for $\Gl=1$
    \\
    ?? & 
    $\left(\log\left(c_1+y\right)\right)^{-\Gl}$ &
    $\text{LinLog}^{(2)}$ &
    $\Gb\rightarrow0$; $c_2=0$
    \\
    Gamma & 
    $y^{-\Gl}e^{-c_1\Gl y}$ &
    $\text{LogLin}^{(1)}$ &
    $\Gb\rightarrow0$; Pearson type III, includes chi-square
    \\
    Gamma-Gauss & 
    $y^{-\Gl}e^{-c_1\Gl y^2}$ &
    $\text{LogLin}^{(1)}$ &
    $\Gb\rightarrow0$; $f_y=y^2$; $m_y=1$ or $m_y=\Tr'$; Rayleigh $\Gl=-1$
    \\
    Generalized gamma & 
    $y^{-\Gg(\Gl-1)-1}e^{-c_1\Gl y^\Gg}$ &
    $\text{LogLin}^{(1)}$ &
    $\Gb\rightarrow0$; $y\rightarrow y^\Gg$; $m_y=y^{\Gg-1}$; Chi for $\Gg=2$ and $c_1\Gl=1/2$
    \\
    Beta & 
    $(c_2-y)^{-\Gl}(y-c_1)^{-b\Gl}$ &
    $\text{LogLinLog}^{(1)}$ &
    $\Gb\rightarrow0$; Pearson type I; $c_1\le y \le c_2$
    \\
    Beta prime/F& 
    $y^{-b\Gl}(1+y)^{(b+1)\Gl-2}$ &
    $\text{LogLinLog}^{(1)}$ &
    $\Gb\rightarrow0$; $y\rightarrow y/(1+y)$; $m_y=(1+y)^{-2}$; $y>0$; Pearson VI
    \\
    Gamma variant& 
    $(c_1+y)^{-b\Gl} e^{-c_2\Gl y}$ &
    $\text{LinLogLin}^{(1)}$ &
    $\Gb\rightarrow0$; $y>0$
    \\
   \hline
  \end{tabular}
  \end{center}
  } %end scriptsize
\begin{flushleft}
\vskip6pt\noindent$^*$Assumptions: base form for $p_y$ is always $m_y e^{-\Gl e^{\Gb w}}$, in which $\Trf=e^{\Gb w}$, as given in \Eq{expscale}. The $w$ column describes the base scale, expressed as combinations of Lin (Linear) and Log scaling, with the superscript denoting the number of recursions as in \Eq{recurse}. For example, $\text{Log}^{(1)}$ implies that $w(f_y)=\log(f_y)$, and $\text{LinLog}^{(1)}$ implies $w(f_y)=\log(c_1+f_y)$. Purely linear scaling is shown as ``Linear,'' which implies $w\equiv f_y$. Recursive expansion of a linear scale remains linear, so no superscript is given for linear scales. Unless otherwise noted, $f_y=y$, shift invariance only is assumed for $\Tr$ with respect to $\GR$ with $\Gb \ne 0$, and $m_y=1$. When $\Gb\rightarrow0$ is shown, affine invariance holds for $\Tr$ with respect to $\GR$. For extreme value distributions, $m_y=\Tr'$ abbreviates the proper change of scale, $m_y=|\Trfp|$, in which information dissipates on the cumulative distribution scale. Change of variable is shown as $y\rightarrow g(y)$, which often leads to a change of scale, $m_y=g'(y)$. Direct values $y$, possibly corrected by displacement from a central location, $y-\Gm$, are shown here as $y$ without correction.  Squared deviations $(y-\Gm)^2$ from a central location are shown here as $y^2$. Listings of distributions can be found in various texts \autocite{johnson94continuous,johnson95continuous,kleiber03statistical}.  Many additional forms can be generated by varying the measurement function.  In the first column, the question marks denote a distribution for which I did not find a commonly used name. Modified from Table 5 of \textcite{frank11a-simple}. See that article for additional details.
\end{flushleft}
\end{table*}

\subsection{A recursive hierarchy for the base scale}

The base scales associated with common distributions typically arise as combinations of linear and logarithmic scaling. For example, the linear-log scale can be defined by $\log(c+x)$. This scale changes linearly in $x$ when $x$ is much smaller than $c$ and logarithmically in $x$ when $x$ is much larger than $c$. As $c\rightarrow0$, the scale becomes almost purely logarithmic, and for large $c$, the scale becomes almost purely linear. 

We can generate a recursive hierarchy of linear-log scale deformations by
\begin{equation}\label{eq:recurse}
  w^{(i)}=\log\left(c_i+w^{(i-1)}\right).
\end{equation}
The hierarchy begins with $w^{(0)}=f_y$, in which $f_y$ denotes our underlying observable. Recursive expansion of the hierarchy yields: a linear scale, $w^{(0)}=f_y$; a linear-log deformation, $w^{(1)}=\log(c_1+f_y)$; a linear-log deformation of the linear-log scale, $w^{(2)}=\log(c_2+\log(c_1 +f_y))$; and so on. A log deformation of a log scale arises as a special case, leading to a double log scale. 

Other scales, such as the log-linear scale, can be expanded in a similarly recursive manner. We may also consider log-linear-log scales and linear-log-linear scales. We can abbreviate a scale, $w$, by its recursive deformation and by its level in a recursive hierarchy. For example, 
\begin{equation}\label{eq:wexample}
 \text{LinLog}^{(2)}=\log(c_2+\log(c_1 +f_y))
\end{equation}
is the second recursive expansion of a linear-log deformation. The initial value for any recursive hierarchy with a superscript of $i=0$ associates with the base observable $w^{(0)}=f_y$, which I will also write as ``Linear,'' because the base observable is always a linear expression of the underlying observable, $f_y$.

\subsection{Examples of common probability distributions}

Table \ref{tab:commonDistn} shows that commonly observed probability distributions arise from combinations of linear and logarithmic scaling. For example, the simple linear-log scale expresses linear scaling at small magnitudes and logarithmic scaling at large magnitudes. The distributions that associate with linear-log scaling include very common patterns. 

For direct observables, $f_y=y$, the linear-log scale includes the purely linear exponential distribution as a limiting case, the purely logarithmic power law (Pareto type I) distribution as a limiting case, and the Lomax (Pareto type II) distribution that is exponential at small magnitudes and has a power law tail at large magnitudes.

For observables that measure the squared distance of fluctuations from a central location, $f_y=(y-\Gm)^2$, or $y^2$ for simplicity, the linear-log scale includes the purely linear Gaussian (normal) distribution as a limiting case, and the generalized Student's distribution that is a Gaussian linear pattern for small deviations from the central location and grades into a logarithmic power law pattern in the tails at large deviations. 

Most of the commonly observed distributions arise from other simple combinations of linear and logarithmic scaling. To mention just two further examples among the many described in Table~\ref{tab:commonDistn}, the log-linear scale leads to the gamma distribution, and the log-linear-log scale leads to the commonly observed beta distribution.

\section{Why do linear and logarithmic scales dominate?}

Processes in the natural world often cause highly nonlinear transformations of inputs into outputs. Why do those complex nonlinear transformations typically lead in the aggregate to simple combinations of linear and logarithmic base scales? Several possibilities exist \autocite{frank13input-output}. I mention a few in this section. However, I do not know of any general answer to this essential question. A clear answer would greatly enhance our understanding of the commonly observed patterns in nature. 

\subsection{Absolute versus relative incremental information}

The scaling of information often changes between linear and logarithmic as magnitude changes. At some magnitudes, a fixed measurement increment provides about the same (linear) information over a varying range, whereas at other magnitudes, a fixed measurement provides less (logarithmic) information as values increase. 

Consider the example of measuring distance \autocite{frank10measurement,frank13input-output}. Start with a ruler that is about the length of your hand. With that ruler, you can measure the size of all the visible objects in your office. That scaling of objects in your office with the length of the ruler means that those objects have a natural linear scaling in relation to your ruler.

Now consider the distances from your office to various galaxies. If the distance is sufficiently great, your ruler is of no use, because you cannot distinguish whether a particular galaxy moves farther away by one ruler unit. Instead, for two distant galaxies, you can measure the ratio of distances from your office to each galaxy. You might, for example, find that one galaxy is twice as far as another, or, in general, that a galaxy is some percentage farther away than another. Percentage changes define a ratio scale of measure, which has natural units in logarithmic measure \autocite{hand04measurement}. For example, a doubling of distance always adds $\log(2)$ to the logarithm of the distance, no matter what the initial distance.

Measurement naturally grades from linear at local magnitudes to logarithmic at distant magnitudes when compared to some local reference scale. The transition between linear and logarithmic varies between problems, depending partly on measurement technology. Measures from some phenomena remain primarily in the linear domain, such as measures of height and weight in humans. Measures for other phenomena remain primarily in the logarithmic domain, such as large cosmological distances. Other phenomena scale between the linear and logarithmic domains, such as fluctuations in the price of financial assets \autocite{aparicio01empirical} or the distribution of income and wealth \autocite{dragulescu01exponential}.

Consider the opposite direction of scaling, from local magnitude to very small magnitude. Your hand-length ruler is of no value for small magnitudes, because it cannot distinguish between a distance that is a fraction $10^{-4}$ of the ruler and a distance that is $2 \times 10^{-4}$ of the ruler. At small distances, one needs a standard unit of measure that is the same order of magnitude as the distinctions to be made. A rule of length $10^{-4}$ distinguishes between $10^{-4}$ and $2 \times 10^{-4}$, but does not distinguish between $10^{-8}$ and $2 \times 10^{-8}$. At small magnitudes, ratios can potentially be distinguished, causing the unit of informative measure to change with scale. Thus, small magnitudes naturally have a logarithmic scaling.

As we change from very small to intermediate to very large, the measurement scaling naturally grades from logarithmic to linear and then again to logarithmic, a log-linear-log scaling \autocite{frank13input-output}. The locus of linearity and the meaning of very small and very large differ between problems, but the overall pattern of the scaling relations remains the same. 

\subsection{Common arithmetic operations lead to common scaling relations}

Perhaps linear and logarithmic scaling reflect aggregation by addition or multiplication of fluctuations. Adding fluctuations often tends in the limit to a smooth linear scaling relation. Multiplying fluctuations often tends in the limit to a smooth logarithmic scaling relation. 

Consider the basic log-linear scale that leads to the gamma distribution. A simple generative model for the gamma distribution arises from the waiting time for the $k$th event to occur. At time zero, no events have occurred. 

At small magnitudes of time, the occurrence of all $k$ events requires essentially simultaneous occurrence of all of those events. Nearly simultaneous occurrence happens roughly in proportion to the product of the probability of any single event occurring in a small time interval. Multiplication associates with logarithmic scaling.

At large magnitudes of time, either all $k$ events have occurred, or in most cases $k-1$ events have occurred and we wait only for the last event. The waiting time for a single event follows an the exponential distribution associated with linear scaling. Thus, the waiting time for $k$ events naturally follows a log-linear pattern. 

Any process that requires simultaneity at extreme magnitudes leads to logarithmic scaling at those limits. Thus, a log-linear-log scale may be a very common underlying pattern. Special cases include log-linear, linear-log, purely log, and purely linear. For those variant patterns, the actual extreme tails may be logarithmic, although difficulty observing the extreme tail pattern may lead to many cases in which a linear tail is a good approximation over the range of observable magnitudes.   

Other aspects of aggregation and limiting processes may also lead to the simple and commonly observed scaling relations. For example, fractal theory provides much insight into logarithmic scaling relations \autocite{mandelbrot83the-fractal,sornette06critical}. However, I do not know of any single approach that matches the simplicity of the commonly observed combinations of linear and logarithmic scaling patterns to a single, simple underlying theory. 

The invariances associated with simple scaling patterns may provide some clues. As noted earlier, shift invariance associates with exponential scaling, and affine invariance associates with linear scaling. It is easy to show that power law invariance associates with logarithmic scaling. For example, in the measurement scale invariance expression given in \Eq{affine}, the invariance holds for a log scale, $\Tr(y)=\log(y)$, in relation to power law transformations of the observables, $\GR(y)=c y^\Gg$, as shown in \Eq{plawinv}. 

We may equivalently say that a scaling relation satisfies power law invariance or that a scaling relation is logarithmic. Noting the invariance does not explain why the scaling relation and the associated invariance are common, but it does provide an alternative and potentially useful way in which to study the problem of commonness. 

\section{Asymptotic invariance}

The measurement functions, $\Tr$, that define maximum entropy distributions satisfy the affine invariance given in \Eq{affine2}, repeated here
\begin{equation}\label{eq:affine2-1}
  \Tr \sim \Tr\circ\GR.
\end{equation}
One can think of $\GR$ as an input-output function that transforms observations in a way that does not change information with respect to probability pattern. 

Most of the commonly observed probability patterns have a simple form, associated with a simple measurement function composed of linear, logarithmic, and exponential components. I have emphasized the open problem of why the measurement functions, $\Tr$, tend to be confined to those simple forms. That simplicity of measurement implies an associated simplicity for the form of $\GR$ under which information remains invariant. If we can figure out why $\GR$ tends to be simple, then perhaps we may understand the simplicity of $\Tr$.

\subsection{Multiple transformations of observations}

At the microscopic scale, observations may tend to get transformed or filtered through a variety of complex processes represented by variable and complex forms of $\GR$. Then, for a simple measurement function, $\Tr$, the fundamental affine invariance would not hold
\begin{equation}\label{eq:notaffine}
  \Tr \not\sim \Tr\circ\GR.
\end{equation}
However, the great lesson of statistical mechanics and maximum entropy is that, for complex underlying processes, aggregation often smooths ultimate pattern into a simple form.  Perhaps multiple filtering of observations through input-output functions $\GR$ would, in the aggregate, lead to a simple overall form for the transformation of initial observations into the actual values observed \autocite{frank13input-output}. 

We can study how multiple applications of input-output transformations may influence the measurement function, $\Tr$. Note that in the basic invariance of \Eq{affine2-1}, application of $\GR$ does not change the information in observations. Thus, we can apply $\GR$ multiple times and still maintain invariant information. If we write $\GR^n$ or $\GR^s$ for $n$ or $s$ applications of input-output processing for $n,s=,0,1,2,\ldots$, then we can write the more general expression for the fundamental measurement and information invariance as
\begin{equation}\label{eq:affine3}
  \Tr\circ\GR^n \sim \Tr\circ\GR^s.
\end{equation}

\subsection{Invariance in the limit}

Suppose that, for a simple measurement function, $\Tr$, and a complex input-output process, $\GR$, the basic invariance does not hold \Eqp{notaffine}. However, it may be that multiple rounds of processing by $\GR$ ultimately lead to a relatively simple transformation of the initial inputs to the final outputs. In other words, $\GR$ may be complex, but for sufficient large $n$, the form of $\GR^n$ may be simple \autocite{frank13input-output}. This aggregate simplicity may lead in the limit to asymptotic invariance
\begin{equation}\label{eq:affinelimit}
  \Tr\circ\GR^n \rightarrow \Tr\circ\GR^\infty
\end{equation}
as $n$ becomes sufficiently large. It is not necessary for every $\GR$ to be identical. Instead, each $\GR$ may be a sample from a pool of alternative transformations. Each individual transformation may be complicated. But in the aggregate, the overall relation between the initial inputs and final outputs may smooth asymptotically into a simple form, such as a power law. If so, then the associated measurement scale smooths asymptotically into a simple logarithmic relation.

Other aggregates of input-output processing may smooth into affine or shift transformations, which associate with linear or exponential scales. When different invariances hold at different magnitudes of the initial inputs, then the measurement scale will change with magnitude. For example, a log-linear scale may reflect asymptotic power law and affine invariances at small and large magnitudes.

%\nsrpart{Discussion}\medskip
\section{Discussion}

Aggregation smooths underlying complexity into simple patterns. The common probability patterns arise by the dissipation of information in aggregates. Each additional random perturbation increases entropy until the distribution of observations takes on the maximum entropy form. That form has lost all information except the constraints on simple average values. 

For each particular probability distribution, the constraint on average value arises on a characteristic measurement scale. That scaling relation, $\Tr$, defines the form of the maximum entropy probability distributions
\begin{equation*}
 p_y \propto m_ye^{-\Gl \Trf}
\end{equation*}
as initially presented in \Eq{general3}, for which $\Tr\equiv\Trf$. Here, $m_y$ accounts for cases in which information dissipates on one scale, but we measure probability pattern on a different scale.

The common probability distributions tend to have simple forms for $\Tr$ that follow linear, logarithmic, or exponential scaling at different magnitudes. The way in which those three fundamental scalings grade into each other as magnitude changes sets the overall scaling relation.

A scaling relation defines the associated maximum entropy distribution. Thus, reading a probability distribution as a statement about process reduces to reading the embedded scaling relation, and trying to understand the processes that cause such scaling. Similarly, understanding the familial relations between probability patterns reduces to understanding the familial relations between different measurements scales. 

The greatest open puzzle concerns why a small number of simple measurement scales dominant the commonly observed patterns of nature. I suggested that the solution may follow from the basic invariance that defines a measurement scale. \EEq{affine2} presented that invariance as
\begin{equation*}
  \Tr \sim \Tr\circ\GR.
\end{equation*}
The measurement scale, $\Tr$, is affine invariant to transformation of the observations by $\GR$. In other words, the information in measurements with regard to probability pattern does not change if we use the directly measured observations or we measure the observations after transformation by $\GR$, when analyzed on the scale $\Tr$. 

In many cases, the small scale processes, $\GR$, that transform underlying values may have complex forms. If so, then the associated scaling relation $\Tr$, might also be complex, leaving open the puzzle of why observable forms of $\Tr$ tend to be simple. I suggested that underlying values may often be transformed by multiple processes before ultimate measurement. Those aggregate transformations may smooth into a simple form with regard to the relation between initial inputs and final measurable outputs. If we express a sequence of $n$ transformations as $\GR^n$, then the asymptotic invariance of the aggregate processing may be simple in the sense that 
\begin{equation}
  \Tr\circ\GR^n \rightarrow \Tr\circ\GR^\infty
\end{equation}
as given by \Eq{affinelimit}. Here, the measurement scaling $\Tr$, and the aggregate input-output processing $\GR^n$ are relatively simple and consistent with commonly observed patterns. 

The puzzle concerns how aggregate input-output processing smooths into simple forms \autocite{frank13input-output}. In particular, how does a combination of transformations lead in the aggregate to a simple asymptotic invariance? 

The scaling pattern for any aggregate input-output relation may have simple asymptotic properties. The application to probability patterns arises when we embed a simple asymptotic scaling relation into the maximum entropy process of dissipating information. The dissipation of information in maximum entropy occurs as measurements are made on the aggregation of individual outputs.

Two particularly simple forms of invariance by $\Tr$ to input-output processing by $\GR^n$ may be important. If $\GR^n$ is a shift transformation $w\mapsto \Gd+w$ for some base scaling $w$, then the associated measurement scale has the form $\Trf=e^{\Gb w}$. This exponential scaling corresponds to the fact that exponential growth or decay is shift invariant. With exponential scaling, the general maximum entropy form is
\begin{equation*}
 p_y \propto m_ye^{-\Gl e^{\Gb w}}.
\end{equation*}
The extreme value distributions and other common distributions derive from that double exponential form. The particular distribution depends on the base scaling, $w$, as illustrated in Table~\ref{tab:commonDistn}. 

Shift transformation is a special case of the broader class of affine transformations, $w\mapsto \Gd+\Gth w$. If $\GR^n$ causes affine changes, then the broader class of input-output relations leads to a narrower range of potential measurement scales that preserve invariance. In particular, an affine measurement scale is the only scale that preserves information about probability pattern in relation to affine transformations. For maximum entropy probability distributions, we may write $\Trf=w$ for the measurement scale that preserves invariance to affine $\GR^n$, leading to the simpler form for probability distributions 
\begin{equation*}
 p_y \propto m_ye^{-\Gl w},
\end{equation*}
which includes most of the very common probability distributions. Thus, the distinction between asymptotic shift and affine changes of initial base scales before potential measurement may influence the general form of probability patterns.

In summary, the common patterns of nature follow a few generic forms. Those forms arise by the dissipation of information and the scaling relations of measurement. The measurement scales arise from the particular way in which the information in a probability pattern is invariant to transformation. Information invariance apparently limits the common measurement scales to simple combinations of linear, logarithmic, and exponential components. Common probability distributions express how those component scales grade into one another as magnitude changes.

%\nsrpart{Methods}

\section{Appendix: scale transformation}

In some cases, information dissipates on one scale, but we wish to express the probability pattern on another scale. For example, a process may lead to a final measured value that is the product of a series of underlying processes. The product of multiple values is equal to the sum of the logarithms of those values. So we may consider how information dissipates as the logarithm of each individual component is added to the total. The theory for the dissipation of information has a particularly simple interpretation as the sum of independent random processes.  

The sum of random processes often converges to a Gaussian distribution, preserving information only about the average squared distance of fluctuations around the mean. Thus, we obtain a simple expression for the dissipation of information when we transform the final measured values, which arise by multiplication, to the additive logarithmic scale. After finding the shape of the distribution on the log transformed scale, it makes sense to transform the distribution of values back to the original scale of the measurements. In the case of a Gaussian distribution on the altered scale, the transformation back to the original scale leads to the pattern known as the log-normal distribution.

The transformations associated with the log-normal distribution are well known. Because the Gaussian distribution is a standard component of simple maximum entropy approaches, the log-normal also falls within that scope. But the Gaussian and log-normal transformation pair are sometimes considered to be a special case. Here, I emphasize that one must understand the structure of the transformation argument more generally. Information often dissipates on one scale, but we may wish to express probability patterns on another scale. 

Once one recognizes the more general structure for the dissipation of information, many previously puzzling patterns fall naturally within the scope of a simple theory of probability patterns. In the main text, I discuss important examples, particularly the extreme value distributions that play a central role in many applications of risk analysis. In this Methods section, I give the general form by which one can express the different scales for the dissipation of information and for measurement. That general form provides the key to reading the mathematical expressions of probability patterns as simple statements about process.

For continuous variables, probability expressions describe the chance that an observation is close to a value $y$. The chance of observing a value exactly equal to $y$ must be close to zero, because there are essentially an infinite number of possible values that $y$ can take on. So we describe probability in terms of the chance that an observation falls into a small interval between $y$ and $y+\dd y$, where $\dd y$ is a small increment. We write the probability of falling into a small increment near $y$ as $p_y|\dd y|$. 

We are interested in understanding the distribution on the scale $y$. But suppose that information dissipates on a different scale given by $x$, leading to the distribution $p_x|\dd x|$. After obtaining the distribution on the scale $x$ by applying the theory for the dissipation of information and constraint, we often wish to transform the distribution to the original scale $y$. The relation between $x$ and $y$ is given by the transformation $x=g(y)$, where $g$ is some function of $y$. For example, we may have $x=\log(y)$. In general, we can use any transformation that has meaning for a particular problem. 

By standard calculus, we can write $\dd x = g'(y)\dd y$, where $g'$ is the derivative of $g$ with respect to $y$. Define $m_y=|g'(y)|$, which gives a notation $m_y$ that emphasizes the term as the translation between the measurement scales for $x$ and $y$. Thus
\begin{equation*}
  |\dd x| = m_y|\dd y|.
\end{equation*}
Because $x=g(y)$, we can also write $p_x=p_{g(y)}$, and so 
\begin{equation*}
  p_x|\dd x| = m_yp_{g(y)}|\dd y|=p_y|\dd y|,
\end{equation*}
or
\begin{equation}\label{eq:changemeasure}
  p_y=m_yp_x=m_yp_{g(y)}.
\end{equation}

Because information dissipates on the scale $x$, we can often find the distribution $p_x$ relatively easily. From \Eq{general1}, the form of that distribution is
\begin{equation*}
 p_x \propto e^{-\Gl f_x}.
\end{equation*}
Applying the change in measure in \Eq{changemeasure}, we obtain
\begin{equation}\label{eq:general3x}
 p_y \propto m_ye^{-\Gl f_{g(y)}}.
\end{equation}
To illustrate, consider the log-normal example, in which $x=g(y)=\log(y)$ and $m_y = y^{-1}$. On the logarithmic scale, $x$, the distribution is Gaussian
\begin{equation*}
 p_x \propto e^{-\Gl (x-\Gm)^2},
\end{equation*}
in which $\Gl=1/2\Gs^2$. From \Eq{changemeasure}, we obtain the distribution on the original scale, $y$, as 
\begin{equation*}
 p_y \propto y^{-1}e^{-\Gl (\log(y)-\Gm)^2},
\end{equation*}
which is the log-normal distribution. The relation between the Gaussian and the log-normal is widely known. But the general principle of studying the dissipation of information on one scale and then transforming to another scale is more general. That relation is an essential step in reading probability expressions in terms of process and in unifying the commonly observed distributions into a single general framework.

\section*{Acknowledgments}

\noindent National Science Foundation grant DEB--1251035 supports my research.

\bigskip
\bibliography{main}

\end{document}